\documentclass[runningheads]{svmult}

\usepackage{makeidx}   
\usepackage{graphicx}  
\usepackage{subeqnar}  
\usepackage{multicol}  
\usepackage{physprbb}  
\makeindex             

\newcommand{\greeksym}[1]{{\usefont{U}{psy}{m}{n}#1}}

\usepackage{amsmath2000}   
\usepackage{url}       
\newcommand{\uTheta}{\mbox{\greeksym{Q}}} 
\usepackage{amssymb}   
\begin{document}
%
%
\begin{titlepage}
\vspace*{3truecm}
\begin{center}
\begin{sffamily}
{\Large Path Integrals and Their Application to\\[0.3\baselineskip]
Dissipative Quantum Systems}\\[2\baselineskip]
{\large Gert-Ludwig Ingold}\\[0.3\baselineskip]
Institut f{\"u}r Physik, Universit{\"a}t Augsburg, D-86135
Augsburg\\[25\baselineskip]
\end{sffamily}
to be published in ``Coherent Evolution in Noisy Environments'',\\
Lecture Notes in Physics, \url{http://link.springer.de/series/lnpp/}\\
\copyright\ Springer Verlag, Berlin-Heidelberg-New York
\end{center}
\end{titlepage}
%
%
%
\title*{Path Integrals and Their Application to Dissipative Quantum Systems}
\toctitle{Path Integrals and Their Application to Dissipative Quantum
Systems}
%
%
\titlerunning{Path Integrals and Quantum Dissipation}
%
\author{Gert-Ludwig Ingold}
\authorrunning{Gert-Ludwig Ingold}
%
%
\institute{Institut f{\"u}r Physik, Universit{\"a}t Augsburg,
D-86135 Augsburg, Germany}

\maketitle              


\setcounter{page}{1}
\section{Introduction}
The coupling of a system to its environment is a recurrent subject in this
collection of lecture notes. The consequences of such a coupling are threefold.
First of all, energy may irreversibly be transferred from the system to the
environment thereby giving rise to the phenomenon of \emph{dissipation}. In 
addition, the fluctuating force exerted by the environment on the system causes 
\emph{fluctuations} of the system degree of freedom which manifest itself for 
example as Brownian motion. While these two effects occur both for classical as 
well as quantum systems, there exists a third phenomenon which is specific to 
the quantum world. As a consequence of the entanglement between system and 
environmental degrees of freedom a coherent superposition of quantum states may 
be destroyed in a process referred to as \emph{decoherence}. This effect is of 
major concern if one wants to implement a quantum computer. Therefore, 
decoherence is discussed in detail in Chap.~5.

Quantum computation, however, is by no means the only topic where the 
coupling to an environment is relevant. In fact, virtually no real system can 
be considered as completely isolated from its surroundings. Therefore, the 
phenomena listed in the previous paragraph play a role in many areas of physics 
and chemistry and a series of methods has been developed to address this 
situation. Some approaches like the master equations discussed in Chap.~2 
are particularly well suited if the coupling to the 
environment is weak, a situation desired in quantum computing. On the other 
hand, in many solid state systems, the environmental coupling can be so strong 
that weak coupling theories are no longer valid. This is the regime where the 
path integral approach has proven to be very useful. 

It would be beyond the scope of this chapter even to attempt to give a 
complete overview of the use of path integrals in the description of
dissipative quantum systems. In particular for a two-level system coupled
to harmonic oscillator degrees of freedom, the so-called spin-boson model,
quite a number of approximations have been developed which are useful in
their respective parameter regimes. This chapter rather attempts to give
an introduction to path integrals for readers unfamiliar with but interested
in this method and its application to dissipative quantum systems. 

In this spirit, Sect.~\ref{gli:sec:pathintegrals} gives an introduction
to path integrals. Some aspects discussed in this section are not necessarily
closely related to the problem of dissipative systems. They rather serve to 
illustrate the path integral approach and to convey to the reader the beauty 
and power of this approach. In Sect.~\ref{gli:sec:dissipation} we elaborate on 
the general idea of the coupling of a system to an environment. The path 
integral formalism is employed to eliminate the environmental degrees of 
freedom and thus to obtain an effective description of the system degree of 
freedom. The results provide the basis for a discussion of the damped harmonic 
oscillator in Sect.~\ref{gli:sec:dho}. Starting from the partition function we 
will examine several aspects of this dissipative quantum system.

Readers interested in a more in-depth treatment of the subject of quantum 
dissipation are referred to existing textbooks. In particular, we recommend the 
book by U.~Weiss \cite{gli:weiss99} which provides an extensive presentation of 
this topic together with a comprehensive list of references. Chapter~4 of 
\cite{gli:dittr98} may serve as a more concise introduction complementary 
to the present chapter. Path integrals are discussed in a whole variety of 
textbooks with an emphasis either on the physical or the mathematical aspects.
We only mention the book by H.~Kleinert \cite{gli:klein95} which gives a
detailed discussion of path integrals and their applications in different areas.

\section{Path Integrals}
\label{gli:sec:pathintegrals}

\subsection{Introduction}
The most often used and taught approach to nonrelativistic quantum mechanics is
based on the Schr{\"o}dinger equation which possesses strong ties with the
the Hamiltonian formulation of classical mechanics. The nonvanishing Poisson 
brackets between position and momentum in classical mechanics lead us to 
introduce noncommuting operators in quantum mechanics. The Hamilton function 
turns into the Hamilton operator, the central object in the Schr{\"o}dinger 
equation. One of the most important tasks is to find the eigenfunctions of the 
Hamilton operator and the associated eigenvalues. Decomposition of a state into 
these eigenfunctions then allows us to determine its time evolution. 

As an alternative, there exists a formulation of quantum mechanics based on
the Lagrange formalism of classical mechanics with the action as the central 
concept. This approach, which was developed by Feynman in the 1940's 
\cite{gli:feynm48,gli:derbe96}, avoids the use of operators though this does 
not necessarily mean that the solution of quantum mechanical problems becomes 
simpler. Instead of finding eigenfunctions of a Hamiltonian one now has to 
evaluate a functional integral which directly yields the propagator required to 
determine the dynamics of a quantum system. Since the relation between 
Feynman's formulation and classical mechanics is very close, the path integral 
formalism often has the important advantage of providing a much more intuitive 
approach as we will try to convey to the reader in the following sections.

\subsection{Propagator}

In quantum mechanics, one often needs to determine the solution 
$\vert\psi(t)\rangle$ of the time-dependent Schr{\"o}dinger equation
\begin{equation}
\I \hbar\frac{\partial \vert\psi\rangle}{\partial t} = H\vert\psi\rangle\;,
\label{gli:eq:schroedinger}
\end{equation}
where $H$ is the Hamiltonian describing the system. Formally, the solution
of (\ref{gli:eq:schroedinger}) may be written as
\begin{equation}
\vert\psi(t)\rangle = {\cal T}\exp\!\left(-\frac\I {\hbar}\int_0^t
\D t'H(t')\right)\vert\psi(0)\rangle\;.
\label{gli:eq:psit}
\end{equation}
Here, the time ordering operator ${\cal T}$ is required because the operators 
corresponding to the Hamiltonian at different times in general due not commute.
In the following, we will restrict ourselves to time-independent Hamiltonians 
where (\ref{gli:eq:psit}) simplifies to
\begin{equation}
\vert\psi(t)\rangle=\exp\!\left(-\frac\I{\hbar}Ht\right)\vert\psi(0)\rangle\;.
\label{gli:eq:psiti}
\end{equation}
As the inspection of (\ref{gli:eq:psit}) and (\ref{gli:eq:psiti}) 
demonstrates, the solution of the time-dependent Schr{\"o}dinger equation 
contains two parts: the initial state $\vert\psi(0)\rangle$ which serves as an 
initial condition and the so-called propagator, an operator which contains all 
information required to determine the time evolution of the system. 

Writing (\ref{gli:eq:psiti}) in position representation one finds
\begin{equation}
\langle x\vert\psi(t)\rangle = \int\!\D x'\langle x\vert\exp\!\left(
-\frac\I {\hbar}Ht\right)\vert x'\rangle \langle x'\vert\psi(0)\rangle
\end{equation}
or
\begin{equation}
\psi(x,t) = \int\!\D x' K(x,t,x',0)\psi(x',0)
\label{gli:eq:psik}
\end{equation}
with the propagator\index{propagator}
\begin{equation}
K(x,t,x',0) = \langle x\vert\exp\!\left(-\frac\I {\hbar}Ht\right)\vert x'
\rangle\;.
\label{gli:eq:propagator}
\end{equation}
It is precisely this propagator which is the central object of Feynman's
formulation of quantum mechanics. Before discussing the path integral
representation of the propagator, it is therefore useful to take a look
at some properties of the propagator.

\begin{figure}[t]
\begin{center}
\includegraphics[width=0.4\textwidth]{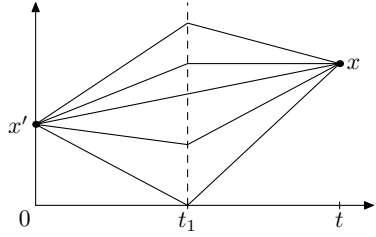}
\end{center}
\caption{According to the semigroup property (\protect\ref{gli:eq:semigroup})
the propagator $K(x,t,x',0)$ may be decomposed into propagators arriving at
some time $t_1$ at an intermediate point $x''$ and propagators continuing
from there to the final point $x$}
\label{gli:fig:semigroup}
\end{figure}

Instead of performing the time evolution of the state $\vert\psi(0)\rangle$
into $\vert\psi(t)\rangle$ in one step as was done in equation 
(\ref{gli:eq:psiti}), one could envisage to perform this procedure in two steps 
by first propagating the initial state $\vert\psi(0)\rangle$ up to an 
intermediate time $t_1$ and taking the new state $\vert\psi(t_1)\rangle$ as 
initial state for a propagation over the time $t-t_1$. This amounts to 
replacing (\ref{gli:eq:psiti}) by
\begin{equation}
\vert\psi(t)\rangle = \exp\!\left(-\frac\I {\hbar}H(t-t_1)\right)
\exp\!\left(-\frac\I {\hbar}Ht_1\right)\vert\psi(0)\rangle
\end{equation}
or equivalently
\begin{equation}
\psi(x,t) = \int\!\D x'\!\int\!\D x'' K(x,t,x'',t_1)K(x'',t_1,x',0)
\psi(x',0)\;.
\label{gli:eq:psik2}
\end{equation}
Comparing (\ref{gli:eq:psik}) and (\ref{gli:eq:psik2}), we find the
semigroup property of the 
propagator\index{propagator!semigroup property of}\index{semigroup property of
propagator}
\begin{equation}
K(x,t,x',0) = \int\!\D x''K(x,t,x'',t_1)K(x'',t_1,x',0)\;.
\label{gli:eq:semigroup}
\end{equation}
This result is visualized in Fig.~\ref{gli:fig:semigroup} where the 
propagators between space-time points are depicted by straight lines
connecting the corresponding two points. At the intermediate time $t_1$ one has 
to integrate over all positions $x''$. This insight will be of use when we 
discuss the path integral representation of the propagator later on.

The propagator contains the complete information about the eigenenergies $E_n$ 
and the corresponding eigenstates $\vert n\rangle$. Making use of the 
completeness of the eigenstates, one finds from (\ref{gli:eq:propagator})
\begin{equation}
K(x,t,x',0) =\sum_n\exp\!\left(-\frac\I {\hbar}E_nt\right)\psi_n(x)
\psi_n(x')^*\;.
\label{gli:eq:kdecomp}
\end{equation}
Here, the star denotes complex conjugation. Not only does the propagator
contain the eigenenergies and eigenstates, this information may also be
extracted from it. To this end, we introduce the retarded Green function
\begin{equation}
G_{\rm r}(x,t,x',0) = K(x,t,x',0)\uTheta(t)    
\end{equation}
where $\uTheta(t)$ is the Heaviside function which equals 1 for positive 
argument $t$ and is zero otherwise. Performing a Fourier transformation, 
one ends up with the spectral representation\index{Green function!spectral 
representation of}\index{spectral representation of Green function}
\begin{equation}
\begin{split}
G_{\rm r}(x,x',E) &= -\frac\I {\hbar}\int_0^{\infty}\D t\exp\!\left(
\frac\I {\hbar} Et\right)G_{\rm r}(t)\\
&= \sum_n\frac{\psi_n(x)\psi_n(x')^*}{E-E_n+\I \varepsilon}\;,
\end{split}
\label{gli:eq:gspek}
\end{equation}
where $\varepsilon$ is an infinitely small positive quantity. According to
(\ref{gli:eq:gspek}), the poles of the energy-dependent retarded
Green function indicate the eigenenergies while the corresponding residua
can be factorized into the eigenfunctions at positions $x$ and $x'$.

\subsection{Free Particle}
\label{gli:subsec:freeparticle}

An important step towards the path integral formulation of quantum mechanics 
can be made by considering the propagator of a free particle of mass $m$.
The eigenstates of the corresponding Hamiltonian
\begin{equation}
H=\frac{p^2}{2m}
\end{equation}
are momentum eigenstates 
\begin{equation}
\psi_p(x)=\frac{1}{\sqrt{2\pi\hbar}}\exp\!\left(\frac\I {\hbar}px\right)
\end{equation}
with a momentum eigenvalue $p$ out of a continuous spectrum. Inserting these 
eigenstates into the representation (\ref{gli:eq:kdecomp}) of the propagator, 
one finds by virtue of
\begin{equation}
\int_{-\infty}^{\infty}\D x\exp(-\I ax^2) 
= \sqrt{\frac{\pi}{\I a}} = \sqrt{\frac{\pi}{a}}\exp\!\left(-{\rm i}
\frac{\pi}{4}\right)
\end{equation}
for the propagator of the free particle\index{propagator!of free particle} 
the result
\begin{equation}
\begin{split}
K(x_{\rm f},t,x_{\rm i} ,0) &= \frac{1}{2\pi\hbar}\int\!\D p
\exp\!\left(-\frac\I {\hbar}\frac{p^2}{2m}t\right)
\exp\!\left(\frac\I {\hbar}p(x_{\rm f}-x_{\rm i})\right)\\
&=\sqrt{\frac{m}{2\pi\I \hbar t}}\exp\!\left(\frac{\rm i}{\hbar}
\frac{m(x_{\rm f}-x_{\rm i})^2}{2t}\right)\;.
\end{split}
\label{gli:eq:fpprop}
\end{equation}

It was already noted by Dirac \cite{gli:dirac33} that the quantum mechanical 
propagator and the classical properties of a free particle are closely related. 
In order to demonstrate this, we evaluate the action of a particle moving from 
$x_{\rm i}$ to $x_{\rm f}$ in time $t$. From the classical path 
\begin{equation}
x_{\rm cl}(s) = x_{\rm i}  + (x_{\rm f}-x_{\rm i})\frac{s}{t}
\end{equation}
obeying the boundary conditions $x_{\rm cl}(0)=x_{\rm i}$ and 
$x_{\rm cl}(t)=x_{\rm f}$, the corresponding classical action is found as
\begin{equation}
S_{\rm cl} = \frac{m}{2}\int_0^t\D s{\dot x}_{\rm cl}^2 =\frac{m}{2}
\frac{(x_{\rm f}-x_{\rm i})^2}{t}\;.
\end{equation}
This result enables us to express the propagator of a free particle entirely
in terms of the classical action as
\begin{equation}
K(x_{\rm f},t,x_{\rm i},0) = \left(-\frac{1}{2\pi{\rm i}\hbar}
\frac{\partial^2S_{\rm cl}(x_{\rm f},t,x_{\rm i},0)}{\partial x_{\rm f}
\partial x_{\rm i}}\right)^{1/2}\exp\!\left(\frac{\rm i}{\hbar}
S_{\rm cl}(x_{\rm f},t,x_{\rm i},0)\right)\;.
\label{gli:eq:kfpcl}
\end{equation}
This result is quite remarkable and one might suspect that it is due to a
peculiarity of the free particle. However, since the propagation in a general 
potential (in the absence of delta function contributions) may be decomposed 
into a series of short-time propagations of a free particle, the result 
(\ref{gli:eq:kfpcl}) may indeed be employed to construct a representation of 
the propagator where the classical action appears in the exponent. In the 
prefactor, the action appears in the form shown in equation 
(\ref{gli:eq:kfpcl}) only within the semiclassical approximation 
(cf.\ Sect.~\ref{gli:subsec:semiclassics}) or for potentials where this 
approximation turns out to be exact.

\subsection{Path Integral Representation of Quantum Mechanics}
\label{gli:subsec:pathintegral}

While avoiding to go too deeply into the mathematical details, we nevertheless
want to sketch the derivation of the path integral representation of the 
propagator. The main idea is to decompose the time evolution over a finite time 
$t$ into $N$ slices of short time intervals $\Delta t=t/N$ where we will 
eventually take the limit $N\to\infty$. Denoting the operator of the kinetic 
and potential energy by $T$ and $V$, respectively, we thus find
\begin{equation}
\exp\!\left(-\frac\I {\hbar}Ht\right) = \left[\exp\!\left(-\frac{\rm i}{\hbar}
(T+V)\Delta t\right)\right]^N\;.
\label{gli:eq:stp}
\end{equation}
For simplicity, we will assume that the Hamiltonian is time-independent
even though the following derivation may be generalized to the
time-dependent case. We now would like to decompose the short-time
propagator in (\ref{gli:eq:stp}) into a part depending on the kinetic energy
and another part containing the potential energy. However, since the two
operators do not commute, we have to exercise some caution. From an
expansion of the Baker-Hausdorff formula one finds
\begin{equation}
\exp\!\left(-\frac\I {\hbar}(T+V)\Delta t\right) \approx
\exp\!\left(-\frac\I {\hbar}T\Delta t\right) 
\exp\!\left(-\frac\I {\hbar}V\Delta t\right) 
+\frac{1}{\hbar^2}[T,V](\Delta t)^2
\label{gli:eq:baker}
\end{equation}
where terms of order $(\Delta t)^3$ and higher have been neglected. Since
we are interested in the limit $\Delta t\to 0$, we may neglect the
contribution of the commutator and arrive at the Trotter formula\index{Trotter
formula}
\begin{equation}
\exp\!\left(-\frac\I {\hbar}(T+V)t\right)= \lim_{N\to\infty}
\left[U(\Delta t)\right]^N
\end{equation}
with the short time evolution operator
\begin{equation}
U(\Delta t) = \exp\!\left(-\frac\I {\hbar}T\Delta t\right) 
\exp\!\left(-\frac\I {\hbar}V\Delta t\right)\;.
\end{equation}
What we have presented here is, of course, at best a motivation and
certainly does not constitute a mathematical proof. We refer readers
interested in the details of the proof and the conditions under which the
Trotter formula holds to the literature \cite{gli:nelso64}.

In position representation one now obtains for the propagator
\begin{equation}
\begin{aligned}
K(x_{\rm f},t,x_{\rm i},0) = \lim_{N\to\infty} \int_{-\infty}^{\infty}
\left(\prod_{j=1}^{N-1}\D x_j\right)&
\left\langle x_{\rm f}\left\vert U(\Delta t)\right\vert x_{N-1}
\right\rangle\dots\\
&\qquad\qquad\times\left\langle x_1\left\vert U(\Delta t)\right\vert x_{\rm i}
\right\rangle\;.
\label{gli:eq:propdecomp}
\end{aligned}
\end{equation}
Since the potential is diagonal in position representation, one obtains
together with the expression (\ref{gli:eq:fpprop}) for the propagator of the 
free particle for the matrix element
\begin{equation}
\begin{aligned}
\left\langle x_{j+1}\left\vert U(\Delta t)\right\vert x_j\right\rangle &=
\left\langle x_{j+1}\left\vert \exp\!\left(-\frac\I {\hbar}T\Delta t\right)
\right\vert x_j\right\rangle 
\exp\!\left(-\frac\I {\hbar}V(x_j)\Delta t\right)\\
&= \sqrt{\frac{m}{2\pi\I \hbar \Delta t}}\exp\!\left[\frac{\rm i}{\hbar}
\left(\frac{m}{2}\frac{(x_{j+1}-x_j)^2}{\Delta t}-V(x_j)\Delta t\right)
\right]\;.
\end{aligned}
\end{equation}
We thus arrive at our final version of the propagator
\begin{equation}
\begin{aligned}
K(x_{\rm f},t,x_{\rm i},0) &= \lim_{N\to\infty} 
\sqrt{\frac{m}{2\pi\I \hbar \Delta t}}\int_{-\infty}^{\infty}
\left(\prod_{j=1}^{N-1}\D x_j
\sqrt{\frac{m}{2\pi\I \hbar\Delta t}}\right)\\
&\qquad\qquad\times\exp\!\left[\frac\I {\hbar}\sum_{j=0}^{N-1}
\left(\frac{m}{2}
\left(\frac{x_{j+1}-x_j}{\Delta t}\right)^2-V(x_j)\right)\Delta t\right]
\end{aligned}
\label{gli:eq:kdis}
\end{equation}
where $x_0$ and $x_N$ should be identified with $x_{\rm i}$ and $x_{\rm f}$,
respectively. The discretization of the propagator used in this expression
is a consequence of the form (\ref{gli:eq:baker}) of the Baker-Hausdorff
relation. In lowest order in $\Delta t$, we could have used a different
decomposition which would have led to a different discretization of the 
propagator. For a discussion of the mathematical subtleties we refer the
reader to \cite{gli:lango82}.

Remarking that the exponent in (\ref{gli:eq:kdis}) contains a discretized 
version of the action
\begin{equation}
S[x]=\int_0^t\D s \left(\frac{m}{2}\dot x^2-V(x)\right)\;,
\label{gli:eq:action}
\end{equation}
we can write this result in short notation as\index{propagator!path integral 
representation of}\index{path integral representation!of propagator}
\begin{equation}
K(x_{\rm f},t,x_{\rm i},0) = \int{\cal D}x \exp\!\left(\frac{\rm i}{\hbar}
S[x]\right)\;.
\label{gli:eq:pathint}
\end{equation}
The action\index{action} (\ref{gli:eq:action}) is a functional which takes as 
argument a function $x(s)$ and returns a number, the action $S[x]$. The 
integral in (\ref{gli:eq:pathint}) therefore is a functional integral where one 
has to integrate over all functions satisfying the boundary conditions 
$x(0)=x_{\rm i}$ and $x(t)=x_{\rm f}$. Since these functions represent paths, 
one refers to this kind of functional integrals also as path 
integral\index{path integral}.

\begin{figure}[t]
\begin{center}
\includegraphics[width=0.5\textwidth]{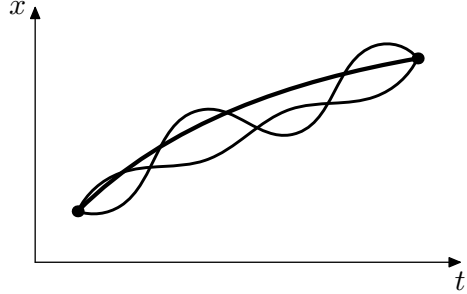}
\end{center}
\caption{The thick line represents a classical path satisfying the boundary
conditions. The thinner lines are no solutions of the classical equation of
motion and may be associated with quantum fluctuations}
\label{gli:fig:paths}
\end{figure}

The three lines shown in Fig.~\ref{gli:fig:paths} represent the infinity
of paths satisfying the boundary conditions. Among them the thicker
line indicates a special path corresponding to an extremum of the action.
According to the principal of least action such a path is a solution of the
classical equation of motion. It should be noted, however, that even
though sometimes there exists a unique extremum, in general there may be more
than one or even none. \label{gli:page:unique} A demonstration of this fact 
will be provided in Sect.~\ref{gli:subsec:dho} where we will discuss the driven 
harmonic oscillator. 

The other paths depicted in Fig.~\ref{gli:fig:paths} may be interpreted as 
quantum fluctuations\index{quantum fluctuations} around the classical path. 
As we will see in Sect.~\ref{gli:subsec:semiclassics}, the amplitude of these 
fluctuations is typically of the order of $\sqrt{\hbar}$. In the classical 
limit $\hbar\to 0$ therefore only the classical paths survive as one should 
expect.

Before explicitly evaluating a path integral, we want to discuss two
examples which will give us some insight into the difference of the approaches
offered by the Schr{\"o}dinger and Feynman formulation of quantum mechanics.

\subsection{Particle on a Ring}
\label{gli:subsec:ring}\index{particle on a ring|(}

We confine a particle of mass $m$ to a ring of radius $R$ and denote its 
angular degree of freedom by $\phi$. This system is described by the 
Hamiltonian
\begin{equation}
H = -\frac{\hbar^2}{2mR^2}\frac{\partial^2}{\partial\phi^2}\;.
\end{equation}
Requiring the wave function to be continuous and differentiable, one finds the 
stationary states
\begin{equation}
\psi_{\ell}(\phi) = \frac{1}{\sqrt{2\pi}}\exp\left(\I \ell\phi\right)
\end{equation}
with $\ell = 0, \pm1, \pm2, \dots$ and the eigenenergies
\begin{equation}
E_{\ell} = \frac{\hbar^2\ell^2}{2mR^2}\;.
\end{equation}
These solutions of the time-independent Schr{\"o}dinger equation allow us to 
construct the propagator\index{propagator!of particle on a ring}
\begin{equation}
K(\phi_{\rm f},t,\phi_{\rm i},0) = \frac{1}{2\pi}\sum_{\ell=-\infty}^{\infty}
\exp\!\left(\I \ell(\phi_{\rm f}-\phi_{\rm i}) - {\rm i}\frac{\hbar\ell^2}
{2mR^2}t\right)\;.
\label{gli:eq:krs}
\end{equation}

\begin{figure}[t]
\begin{center}
\includegraphics[width=0.5\textwidth]{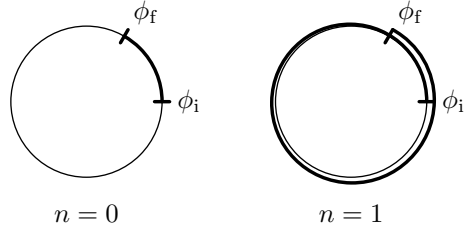}
\end{center}
\caption{On a ring, the angles $\phi_{\rm f}$ and $\phi_{\rm f}+2\pi n$ have
to be identified. As a consequence, there exist infinitely many classical paths 
connecting two points on a ring, which may be identified by their winding
number $n$\index{winding number}}
\label{gli:fig:winding}
\end{figure}

We now want to derive this result within the path integral formalism. To this
end we will employ the propagator of the free particle. However, an important 
difference between a free particle and a particle on a ring deserves our
attention. Due to the ring topology we have to identify all angles $\phi+2\pi 
n$, where $n$ is an integer, with the angle $\phi$. As a consequence, 
there exist infinitely many classical paths connecting $\phi_{\rm i} $ and 
$\phi_{\rm f}$. All these paths are topologically different and can be 
characterized by their winding number $n$\index{winding number}. As an example, 
Fig.~\ref{gli:fig:winding} shows a path for $n=0$ and $n=1$. Due to their 
different topology, these two paths (and any two paths corresponding to 
different winding numbers) cannot be continuously transformed into each other. 
This implies that adding a fluctuation to one of the classical paths will never 
change its winding number. 

Therefore, we have to sum over all winding numbers in order to account for all 
possible paths. The propagator thus consists of a sum over free propagators 
corresponding to different winding numbers\index{propagator!of particle on a
ring}
\begin{equation}
K(\phi_{\rm f},t,\phi_{\rm i},0) = \sum_{n=-\infty}^{\infty}R
\sqrt{\frac{m}{2\pi\I \hbar t}}\exp\!\left(\frac{\rm i}{\hbar}\frac{mR^2}{2}
\frac{(\phi_{\rm f}-\phi_{\rm i}-2\pi n)^2}{t}\right)\;.
\label{gli:eq:krfp}
\end{equation}
Here, the factor $R$ accounts for the fact that, in contrast to the free 
particle, the coordinate is given by an angle instead of a position.

The propagator (\ref{gli:eq:krfp}) is $2\pi$-periodic in $\phi_{\rm f}-
\phi_{\rm i}$ and can therefore be expressed in terms of a Fourier series 
\begin{equation}
K(\phi_{\rm f},t,\phi_{\rm i},0) = \sum_{\ell=-\infty}^{\infty}c_{\ell}
\exp\left[\I \ell(\phi_{\rm f}-\phi_{\rm i})\right]\;.
\end{equation}
The Fourier coefficients are found to read
\begin{equation}
c_{\ell} = \frac{1}{2\pi}\exp\!\left(-\I \frac{\hbar\ell^2}{2mR^2}t\right)
\end{equation}
which proves the equivalence of (\ref{gli:eq:krfp}) with our previous result
(\ref{gli:eq:krs}). We thus have obtained the propagator of a free particle
on a ring both by solving the Schr{\"o}dinger equation and by employing path 
integral methods. These two approaches make use of complementary 
representations. In the first case, this is the angular momentum representation 
while in the second case, one works in the phase representation and sums over 
winding numbers.
\index{particle on a ring|)}

\subsection{Particle in a Box}
\label{gli:subsec:box} \index{particle in a box|(}

Another textbook example in standard quantum mechanics is the particle in a box
of length $L$ confined by infinitely high walls at $x=0$ and $x=L$. From the 
eigenvalues
\begin{equation}
E_j = \frac{\hbar^2\pi^2j^2}{2mL^2}
\label{gli:eq:energybox}
\end{equation}
with $j=1,2,\dots$ and the corresponding eigenfunctions
\begin{equation}
\psi_j(x) = \sqrt{\frac{2}{L}}\sin\left(\pi j\frac{x}{L}\right)
\end{equation}
the propagator is immediately obtained as\index{propagator!of particle in a box}
\begin{equation}
K(x_{\rm f},t,x_{\rm i},0) = \frac{2}{L}\sum_{j=1}^{\infty}
\exp\!\left(-\I \frac{\hbar\pi^2j^2}{2mL^2}t\right)
\sin\!\left(\pi j\frac{x_{\rm f}}{L}\right)
\sin\!\left(\pi j\frac{x_{\rm i}}{L}\right)\;.
\label{gli:eq:propbox}
\end{equation}

\begin{figure}[t]
\begin{center}
\includegraphics[width=0.5\textwidth]{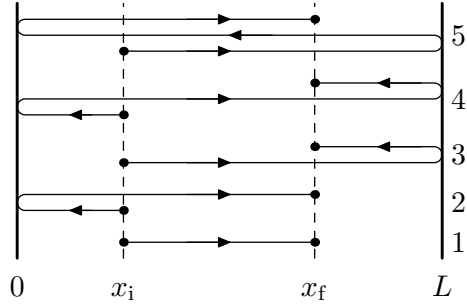}
\end{center}
\caption{The reflection at the walls of a box leads to an infinite number
of possible trajectories connecting two points in the box}
\label{gli:fig:boxclass}
\end{figure}

It took some time until this problem was solved within the path
integral approach \cite{gli:janke79,gli:goodm81}. Here, we have to consider 
all paths connecting the points $x_{\rm i}$ and $x_{\rm f}$ within a period 
of time $t$. Due to the reflecting walls, there again exist infinitely many 
classical paths, five of which are depicted in Fig.~\ref{gli:fig:boxclass}. 
However, in contrast to the case of a particle on a ring, these paths are no 
longer topologically distinct. As a consequence, we may deform a classical path 
continuously to obtain one of the other classical paths. 

\begin{figure}[t]
\begin{center}
\includegraphics[width=0.9\textwidth]{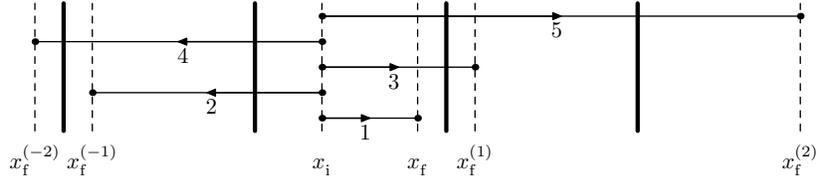}
\end{center}
\caption{Instead of a particle getting reflected at the walls of the box one
may think of a free particle moving from the starting point in the box to
the end point in one of the replicas of the box}
\label{gli:fig:boxfree}
\end{figure}

If, for the moment, we disregard the details of the reflections at the wall, 
the motion of the particle in a box is equivalent to the motion of a free 
particle. The fact that paths are folded back on themselves can be 
accounted for by taking into account replicas of the box as shown in
Fig.~\ref{gli:fig:boxfree}. Now, the path does not necessarily end at 
$x_{\rm f}^{(0)}=x_{\rm f}$ but at one of the mirror images $x_{\rm f}^{(n)}$ 
where $n$ is an arbitrary integer. In order to obtain the propagator, we 
will have to sum over all replicas. Due to the different geometry we need 
to distinguish between those paths arising from an even and an odd number of 
reflections. From Fig.~\ref{gli:fig:boxfree} one can see that for an odd number 
$2n-1$ of reflections, the end point lies at $2nL-x_{\rm f}$ and the 
contribution to the full propagator therefore is given by
\begin{equation}
K^{(2n-1)}(x_{\rm f},t,x_{\rm i},0)=\sqrt{\frac{m}{2\pi\I \hbar t}}
\exp\!\left(\frac{\rm i}{\hbar}
\frac{m(2nL-x_{\rm f}-x_{\rm i})^2}{2t}\right)\;.
\label{gli:eq:kodd}
\end{equation}
On the other hand, for an even number $2n$ of reflections, the end point is
located at $2nL+x_{\rm f}$ and we find
\begin{equation}
K^{(2n)}(x_{\rm f},t,x_{\rm i},0)=\sqrt{\frac{m}{2\pi\I \hbar t}}
\exp\!\left(\frac{\rm i}{\hbar}
\frac{m(2nL+x_{\rm f}-x_{\rm i})^2}{2t}\right)\;.
\label{gli:eq:keven}
\end{equation}
However, it is not obvious that just summing up the propagators 
(\ref{gli:eq:kodd}) and (\ref{gli:eq:keven}) for all $n$ will do the 
job. 

\begin{figure}[t]
\begin{center}
\includegraphics[width=0.5\textwidth]{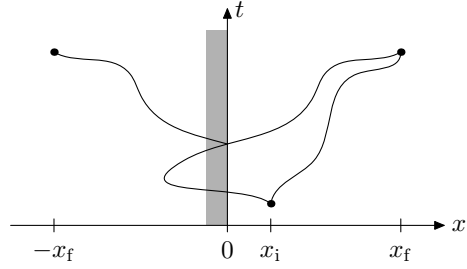}
\end{center}
\caption{A path crossing the wall is cancelled by a path running to the
mirror point of the end point}
\label{gli:fig:wallfluc}
\end{figure}

In order to clarify this point, we start with the somewhat simpler situation
of just one wall and take a look at all paths running between $x_{\rm i}$ and 
$x_{\rm f}$ in time $t$. As can be seen from the space-time diagram in 
Fig.~\ref{gli:fig:wallfluc} there are paths which do not cross the wall and 
which therefore contribute to the path integral. On the other hand, there exist 
also paths which cross the wall an even number of times. Since these paths 
spend some time in the forbidden region, they do not contribute to the path 
integral. 

It requires some thinking to ensure that only paths not crossing the wall are 
taken into account. Our strategy will consist in first writing down a 
propagator $K_{\rm free}$ which disregards the wall. Then, we have to subtract 
off the contributions of all the paths which cross the wall. This can be done 
by constructing a path with the same action as the original path. To this end 
we take the original path up to the last crossing with the wall and then 
continue along the mirror image of the original path. We thus end up at the 
mirror image $-x_{\rm f}$ of the original end point $x_{\rm f}$. Note that a 
path running from $x_{\rm i}$ to $-x_{\rm f}$ necessarily crosses the wall at 
least once. As a consequence, subtracting the propagator between these two 
points eliminates all original paths which do not remain in the region $x>0$. 
We therefore obtain our desired result, the propagator $K_{\rm wall}$ in the 
presence of a wall, by subtracting a propagator going to the reflected end 
point from the unconstrained propagator to the original end point
\cite{gli:janke79,gli:goodm81,gli:auerb97}\index{propagator!in presence of a
wall}
\begin{equation}
K_{\rm wall}(x_{\rm f},t,x_{\rm i},0) = K_{\rm free}(x_{\rm f},t,x_{\rm i},0)
-K_{\rm free}(-x_{\rm f},t,x_{\rm i},0)\;.
\label{gli:eq:wallsubtr}
\end{equation}
This result bears much resemblance with the method of image 
charges\index{method of image charges} in electrostatics. After giving it some 
thought, this should not be too surprising since the free Schr{\"o}dinger 
equation and the Poisson equation are formally equivalent. According to the 
method of image charges one may account for a metallic plate (i.e. the wall) by 
putting a negative charge (i.e. the mirrored end point) complementing the 
positive charge (i.e. the original end point). For the propagator this results 
in the difference appearing in (\ref{gli:eq:wallsubtr}).

Let us now come back to our infinitely deep potential well with two walls.
This problem corresponds to the electrostatics of a charge between two
parallel metal plates. In this case, the method of image charges leads to an 
infinite number of charges of alternating signs. The original positive charge 
gives rise to two negative charges which are each an image corresponding to one 
of the two metal plates.  In addition, however, these images have mirror images 
corresponding to the other metal plate and this process has to be carried on ad 
infinitum. 

Expressing the propagator of the particle in the box in terms of the free
propagator works in exactly the same way. A path intersecting both walls is
subtracted twice, i.e. one time too often. Therefore, one contribution has
to be restored which is done by adding another end point. Continuing the
procedure one ends up with an infinite number of end points, some of which
we have shown in Fig.~\ref{gli:fig:boxfree}. As a consequence, we can
attribute a sign to each end point in this figure. The general rule which
follows from these considerations is that each reflection at a wall leads
to factor $-1$. The propagator therefore can be written as\index{propagator!of
particle in a box}
\begin{equation}
\begin{split}
K(x_{\rm f},t,x_{\rm i},0)=\sqrt{\frac{m}{2\pi{\rm i}\hbar t}}
\sum_{n=-\infty}^{\infty}&\Bigg[\exp\!\left(\frac\I {\hbar}
\frac{m(2nL+x_{\rm f}-x_{\rm i})^2}{2t}\right)\\
&\phantom{\Bigg[}-\exp\!\left(\frac\I {\hbar}
\frac{m(2nL-x_{\rm f}-x_{\rm i})^2}{2t}\right)\Bigg]\;.
\end{split}
\label{gli:eq:propboxf}
\end{equation}
The symmetries 
\begin{align}
K(x_{\rm f}+2L,t,x_{\rm i},0)&=K(x_{\rm f},t,x_{\rm i},0)\\
K(-x_{\rm f},t,x_{\rm i},0)&=-K(x_{\rm f},t,x_{\rm i},0) 
\end{align}
suggest to expand the propagator into the Fourier series
\begin{equation}
K(x_{\rm f},t,x_{\rm i},0) = \sum_{j=1}^{\infty} a_j(x_{\rm i},t)
\sin\left(\pi j\frac{x_{\rm f}}{L}\right)\;.
\label{gli:eq:decompbox}
\end{equation}
Its Fourier coefficients are obtained from (\ref{gli:eq:propboxf}) as
\begin{equation}
\begin{split}
a_j(x_{\rm i},t)&=\frac{1}{L}\int_{-L}^{L}\D x_{\rm f} \sin\left(\pi j
\frac{x_{\rm f}}{L}\right)K(x_{\rm f},t,x_{\rm i},0)\\
&=\frac{2}{L}\sin\left(\pi j\frac{x_{\rm i}}{L}\right)
\exp\!\left(-\frac\I {\hbar}E_jt\right)
\label{gli:eq:coeffbox}
\end{split}
\end{equation}
where the energies $E_j$ are the eigenenergies of the box defined in
(\ref{gli:eq:energybox}). Inserting (\ref{gli:eq:coeffbox}) into
(\ref{gli:eq:decompbox}) we thus recover our previous result
(\ref{gli:eq:propbox}).\index{particle in a box|)}

\subsection{Driven Harmonic Oscillator}
\label{gli:subsec:dho}\index{driven harmonic oscillator|(}

Even though the situations dealt with in the previous two sections have been
conceptually quite interesting, we could in both cases avoid the explicit
calculation of a path integral. In the present section, we will introduce the
basic techniques needed to evaluate path integrals. As an example, we will
consider the driven harmonic oscillator which is simple enough to allow for
an exact solution. In addition, the propagator will be of use in the discussion
of damped quantum systems in later sections.

Our starting point is the Lagrangian
\begin{equation}
L = \frac{m}{2}\dot x^2 - \frac{m}{2}\omega^2x^2 +xf(t)
\label{gli:eq:lagrangian}
\end{equation}
of a harmonic oscillator with mass $m$ and frequency $\omega$. The force $f(t)$
may be due to an external field, e.g. an electric field coupling via dipole
interaction to a charged particle. In the context of dissipative quantum 
mechanics, the harmonic oscillator could represent a degree of freedom of the 
environment under the influence of a force exerted by the system.

According to (\ref{gli:eq:pathint}) we obtain the propagator $K(x_{\rm f},t,
x_{\rm i},0)$ by calculating the action for all possible paths starting at
time zero at $x_{\rm i}$ and ending at time $t$ at $x_{\rm f}$. It is 
convenient to decompose the paths 
\begin{equation}
x(s) = x_{\rm cl}(s)+\xi(s) 
\label{gli:eq:clfluc}
\end{equation}
into the classical path $x_{\rm cl}$ satisfying the boundary conditions 
$x_{\rm cl}(0)=x_{\rm i}$, $x_{\rm cl}(t)=x_{\rm f}$ and a fluctuating part 
$\xi$ vanishing at the boundaries, i.e. $\xi(0)=\xi(t)=0$. The classical path
has to satisfy the equation of motion
\begin{equation}
m\ddot x_{\rm cl} + m\omega^2 x_{\rm cl} = f(s)
\end{equation}
obtained from the Lagrangian (\ref{gli:eq:lagrangian}).

For an exactly solvable problem like the driven harmonic oscillator, we could
replace $x_{\rm cl}$ by any path satisfying $x(0)=x_{\rm i}$, $x(t)=x_{\rm f}$.
We leave it as an exercise to the reader to perform the following calculation
with $x_{\rm cl}(s)$ of the driven harmonic oscillator replaced by $x_{\rm i}+
(x_{\rm f}-x_{\rm i})s/t$. However, it is important to note that within the
semiclassical approximation discussed in Sect.~\ref{gli:subsec:semiclassics}
an expansion around the classical path is essential since this path leads to
the dominant contribution to the path integral.

With (\ref{gli:eq:clfluc}) we obtain for the action
\begin{align}
S&= \int_0^t\D s\left(\frac{m}{2}\dot x^2 -\frac{m}{2}\omega^2x^2 +xf(s)
\right)\nonumber\\
&= \int_0^t\D s\left(\frac{m}{2}\dot x_{\rm cl}^2-\frac{m}{2}\omega^2
x_{\rm cl}^2+x_{\rm cl}f(s)\right)
+\int_0^t\D s\left(m\dot x_{\rm cl}\dot\xi - m\omega^2x_{\rm cl}\xi 
+\xi f(s)\right)\nonumber\\
&\qquad +\int_0^t{\rm
d}s\left(\frac{m}{2}\dot\xi^2-\frac{m}{2}\omega^2\xi^2\right).
\label{gli:eq:sdo}
\end{align}
For our case of a harmonic potential, the third term is independent of the 
boundary values $x_{\rm i}$ and $x_{\rm f}$ as well as of the external driving. 
The second term vanishes as a consequence of the expansion around the classical
path. This can be seen by partial integration and by making use of the fact 
that $x_{\rm cl}$ is a solution of the classical equation of motion:
\begin{equation}
\int_0^t\D s\big(m\dot x_{\rm cl}\dot\xi - m\omega^2x_{\rm cl}\xi +\xi
f(s)\big) = -\int_0^t\D s\big(m\ddot x_{\rm cl} + m\omega^2x_{\rm cl} -
f(s)\big)\xi=0\;.
\label{gli:eq:sdopi}
\end{equation}

We now proceed in two steps by first determining the contribution of the 
classical path and then addressing the fluctuations. The solution of the 
classical equation of motion satisfying the boundary conditions reads
\begin{align}
\label{gli:eq:clsol}
x_{\rm cl}(s) =&\ x_{\rm f}\frac{\sin(\omega s)}{\sin(\omega t)}
+x_{\rm i}\frac{\sin(\omega(t-s))}{\sin(\omega t)}\\
&+ \frac{1}{m\omega}
\left[\int_0^s\D u\sin(\omega(s-u))f(u) - \frac{\sin(\omega s)}{\sin(\omega
t)}\int_0^t\D u \sin(\omega(t-u))f(u)\right].\nonumber
\end{align}
A peculiarity of the harmonic oscillator in the absence of driving is the
appearance of conjugate points\index{conjugate point} at times 
$T_n=(\pi/\omega)n$ where $n$ is an arbitrary integer. Since the frequency of 
the oscillations is independent of the amplitude, the position of the 
oscillator at these times is determined by the initial position: 
$x(T_{2n+1})=-x_{\rm i}$ and $x(T_{2n})=x_{\rm i}$. This also illustrates the 
fact mentioned on p.~\pageref{gli:page:unique}, that depending on the boundary 
conditions there may be more than one or no classical solution.

\begin{figure}[t]
\begin{center}
\includegraphics[width=0.4\textwidth]{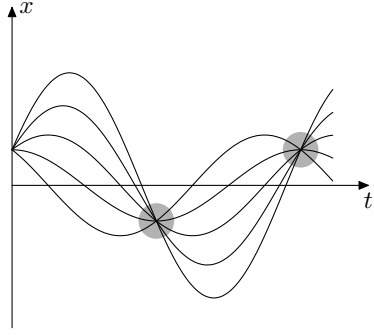}
\end{center}
\caption{In a harmonic potential all trajectories emerging from the same
starting point converge at conjugate points at multiples of half an oscillation
period}
\label{gli:fig:osccaust}
\end{figure}

The task of evaluating the action of the classical path may be simplified
by a partial integration
\begin{equation}
\begin{aligned}
S_{\rm cl}  &= \int_0^t\D s\left(\frac{m}{2}\dot x_{\rm cl}^2 -
\frac{m}{2}\omega^2x_{\rm cl}^2+x_{\rm cl}f(s)\right)\\
&= \left.\frac{m}{2}x_{\rm cl}\dot x_{\rm cl}\right\vert_0^t - \int_0^t\D s
\left(\frac{m}{2}x_{\rm cl}\ddot x_{\rm cl} + \frac{m}{2}\omega^2x_{\rm cl}^2
-x_{\rm cl}f(s)\right)\\
&= \frac{m}{2}\left(x_{\rm f}\dot x_{\rm cl}(t) - x_{\rm i}\dot x_{\rm
cl}(0)\right) + \frac{1}{2}\int_0^t\D s\,x_{\rm cl}(s)f(s)
\end{aligned}
\label{gli:eq:clact}
\end{equation}
where we have made use of the classical equation of motion to obtain the third
line. From the solution (\ref{gli:eq:clsol}) of the classical equation of
motion we get
\begin{align}
\dot x_{\rm cl}(0) &= \omega\frac{x_{\rm f}-x_{\rm
i}\cos(\omega t)}{\sin(\omega t)} - \frac{1}{m\sin(\omega t)}\int_0^t\D s 
\sin(\omega(t-s))f(s)\\
\dot x_{\rm cl}(t) &= \omega\frac{x_{\rm f}\cos(\omega t)
-x_{\rm i}}{\sin(\omega t)} + \frac{1}{m\sin(\omega t)}\int_0^t\D s 
\sin(\omega s)f(s)\;.
\end{align}
Inserting initial and final velocity into (\ref{gli:eq:clact}) we find for the 
classical action
\begin{equation}
\begin{aligned}
S_{\rm cl} =& \frac{m\omega}{2\sin(\omega t)}\left[(x_{\rm i}^2+x_{\rm f}^2)
\cos(\omega t) -2x_{\rm i}x_{\rm f}\right]\\
&+\frac{x_{\rm f}}{\sin(\omega t)}\int_0^t\D s\sin(\omega s)f(s)
+\frac{x_{\rm i}}{\sin(\omega t)}\int_0^t\D s\sin(\omega(t-s))f(s)\\
&-\frac{1}{m\omega\sin(\omega t)}\int_0^t\D s\int_0^s\D u \sin(\omega
u)\sin(\omega(t-s))f(s)f(u)\;.
\label{gli:eq:dhoaction}
\end{aligned}
\end{equation}

As a second step, we have to evaluate the contribution of the fluctuations
which is determined by the third term in (\ref{gli:eq:sdo}). After partial
integration this term becomes
\begin{equation}
S^{(2)}=\int_0^t\D s\left(\frac{m}{2}\dot\xi^2-\frac{m}{2}\omega^2\xi^2
\right)
=-\int_0^t\D s\frac{m}{2}\xi\left(\frac{\D^2}{\D s^2}+\omega^2
\right)\xi\;.
\end{equation}
Here, the superscript `(2)' indicates that this term corresponds to the 
contribution of second order in $\xi$. In view of the right-hand side it is
appropriate to expand the fluctuation
\begin{equation}
\xi(s) =\sum_{n=1}^{\infty}a_n\xi_n(s)
\label{gli:eq:expansion}
\end{equation}
into eigenfunctions of
\begin{equation}
\left(\frac{\D^2}{\D s^2}+\omega^2\right)\xi_n=\lambda_n\xi_n
\label{gli:eq:evdho}
\end{equation}
with $\xi_n(0)=\xi_n(t)=0$. As eigenfunctions of a selfadjoint operator, 
the $\xi_n$ are complete and may be chosen orthonormal. Solving 
(\ref{gli:eq:evdho}) yields the eigenfunctions 
\begin{equation}
\xi_n(s) = \sqrt{\frac{2}{t}}\sin\left(\pi n\frac{s}{t}\right)
\label{gli:eq:efdho}
\end{equation}
and corresponding eigenvalues
\begin{equation}
\lambda_n = -\left(\frac{\pi n}{t}\right)^2+\omega^2\;.
\label{gli:eq:evalho}
\end{equation}
We emphasize that (\ref{gli:eq:expansion}) is not the usual Fourier 
series on an interval of length $t$. Such an expansion could be used in
the form
\begin{equation}
\xi(s) = \sqrt{\frac{2}{t}}\sum_{n=1}^{\infty}\left[a_n\left(
\cos(2\pi n\frac{s}{t})-1\right)+b_n\sin(2\pi n\frac{s}{t})\right]
\label{gli:eq:fourier}
\end{equation}
which ensures that the fluctuations vanish at the boundaries. We invite the 
reader to redo the following calculation with the expansion 
(\ref{gli:eq:fourier}) replacing (\ref{gli:eq:expansion}). While at the end the 
same propagator should be found, it will become clear why the expansion 
in terms of eigenfunctions satisfying (\ref{gli:eq:evdho}) is preferable.

The integration over the fluctuations now becomes an integration over the
expansion coefficients $a_n$. Inserting the expansion (\ref{gli:eq:expansion})
into the action one finds
\begin{equation}
S^{(2)} = -\frac{m}{2}\sum_{n=1}^{\infty}\lambda_na_n^2 = \frac{m}{2}
\sum_{n=1}^{\infty}\left(\left(\frac{\pi n}{t}\right)^2-\omega^2\right)a_n^2\;.
\end{equation}
As this result shows, the classical action is only an extremum of the action
but not necessarily a minimum although this is the case for short time 
intervals $t<\pi/\omega$. The existence of conjugate points at times 
$T_n=n\pi/\omega$ mentioned above manifests itself here as vanishing of the 
eigenvalue $\lambda_n$. Then the action is independent of $a_n$ which implies 
that for a time interval $T_n$ all paths $x_{\rm cl}+a_n\xi_n$ with arbitrary 
coefficient $a_n$ are solutions of the classical equation of motion. 

After expansion of the fluctuations in terms of the eigenfunctions 
(\ref{gli:eq:efdho}), the propagator takes the form
\begin{equation}
K(x_{\rm f},t,x_{\rm i},0)\sim\exp\!\left(\frac{\rm i}{\hbar}S_{\rm cl}\right)
\int\left(\prod_{n=1}^{\infty}\D a_n\right)
\exp\!\left(-\frac\I {\hbar}\frac{m}{2}\sum_{n=1}^{\infty}\lambda_n
a_n^2\right)\;.
\end{equation}
In principle, we need to know the Jacobi determinant of the transformation from
the path integral to the integral over the Fourier coefficients. However, since
this Jacobi determinant is independent of the oscillator frequency $\omega$, 
we may also compare with the free particle. Evaluating the Gaussian fluctuation 
integrals, we find for the ratio between the prefactors of the propagators 
$K_{\omega}$ and $K_0$ of the harmonic oscillator and the free particle, 
respectively,
\begin{equation}
\frac{K_{\omega}\exp[-(\I/\hbar)S_{{\rm cl},\omega}]}
{K_0\exp[-(\I/\hbar)S_{{\rm cl},0}]} = \sqrt{\frac{D_0}{D}}\;.
\label{gli:eq:kratio}
\end{equation}
Here, we have introduced the fluctuation determinants\index{fluctuation 
determinant} for the harmonic 
oscillator
\begin{equation}
D=\det\left(\frac{\D^2}{\D s^2}+\omega^2\right) = \prod_{n=1}^{\infty}
\lambda_n
\label{gli:eq:det}
\end{equation}
and the free particle
\begin{equation}
D_0=\det\left(\frac{\D^2}{\D s^2}\right) = \prod_{n=1}^{\infty}
\lambda_n^0\;.
\label{gli:eq:det0}
\end{equation}
The eigenvalues for the free particle
\begin{equation}
\lambda_n^0=-\left(\frac{\pi n}{t}\right)^2
\end{equation}
are obtained from the eigenvalues (\ref{gli:eq:evalho}) of the harmonic 
oscillator simply by setting the frequency $\omega$ equal to zero. 
With the prefactor of the propagator of the free particle
\begin{equation}
K_0\exp\!\left(-\frac{\I}{\hbar}S_{{\rm cl},0}\right) =
\sqrt{\frac{m}{2\pi\I\hbar t}}
\end{equation}
and (\ref{gli:eq:kratio}), the propagator of the harmonic oscillator becomes
\begin{equation}
K(x_{\rm f},t,x_{\rm i},0) = \sqrt{\frac{m}{2\pi{\rm i}\hbar t}}
\sqrt{\frac{D_0}{D}}\exp\!\left(\frac\I {\hbar}S_{\rm cl}\right)\;.
\label{gli:eq:prophoi}
\end{equation}

For readers unfamiliar with the concept of determinants of differential 
operators we mention that we may define matrix elements of an operator by
projection onto a basis as is familiar from standard quantum mechanics. The 
operator represented in its eigenbasis yields a diagonal matrix with the 
eigenvalues on the diagonal. Then, as for finite dimensional matrices, the 
determinant is the product of these eigenvalues.

Each of the determinants (\ref{gli:eq:det}) and (\ref{gli:eq:det0}) by itself 
diverges. However, we are interested in the ratio between them which is 
well-defined \cite{gli:grads80}
\begin{equation}
\frac{D}{D_0}=\prod_{n=1}^{\infty}\left(1-\left(\frac{\omega t}{\pi n}\right)^2
\right)=\frac{\sin(\omega t)}{\omega t}\;.
\end{equation}
Inserting this result into (\ref{gli:eq:prophoi}) leads to the propagator of 
the driven harmonic oscillator in its final form\index{propagator!of driven
harmonic oscillator}
\begin{equation}
\begin{aligned}
K(x_{\rm f},t,x_{\rm i},0) & = \sqrt{\frac{m\omega}{2\pi{\rm i}\hbar\sin(\omega
t)}}\exp\!\left[\frac\I {\hbar}S_{\rm cl}\right]\\
& = \sqrt{\frac{m\omega}{2\pi\hbar
\vert\sin(\omega t)\vert}}\exp\!\left[\frac\I {\hbar}S_{\rm cl}-\I
\left(\frac{\pi}{4}+n\frac{\pi}{2}\right)\right]
\end{aligned}
\label{gli:eq:propdho}
\end{equation}
with the classical action defined in (\ref{gli:eq:dhoaction}). The Morse 
index\index{Morse index} $n$ in the phase factor is given by the integer part 
of $\omega t/\pi$. This phase accounts for the changes in sign of the sine 
function \cite{gli:morse66}. Here, one might argue that it is not obvious which 
sign of the square root one has to take. However, the semigroup property 
(\ref{gli:eq:semigroup}) allows to construct propagators across conjugate 
points\index{conjugate point} by joining propagators for shorter time 
intervals. In this way, the sign may be determined unambiguously 
\cite{gli:horva79}.

It is interesting to note that the phase factor $\exp(-\I n\pi/2)$ in 
(\ref{gli:eq:propdho}) implies that $K(x_{\rm f},2\pi/\omega,x_{\rm i},0)
= -K(x_{\rm f},0,x_{\rm i},0) = -\delta(x_{\rm f}-x_{\rm i})$, i.e. the wave
function after one period of oscillation differs from the original wave 
function by a factor $-1$. The oscillator thus returns to its original state
only after two periods very much like a spin-1/2 particle which picks up a 
sign under rotation by $2\pi$ and returns to its original state only after a
$4\pi$-rotation. This effect might be observed in the case of the harmonic 
oscillator by letting interfere the wave functions of two oscillators with 
different frequency \cite{gli:rohrl88}.

\index{driven harmonic oscillator|)}

\subsection{Semiclassical Approximation}
\label{gli:subsec:semiclassics}

The systems considered so far have been special in the sense that an exact
expression for the propagator could be obtained. This is a consequence of the
fact that the potential was at most quadratic in the coordinate. 
Unfortunately, in most cases of interest the potential is more complicated and 
apart from a few exceptions an exact evaluation of the path integral turns out
to be impossible. To cope with such situations, approximation schemes have been 
devised. In the following, we will restrict ourselves to the most important 
approximation which is valid whenever the quantum fluctuations are small or, 
equivalently, when the actions involved are large compared to Planck's constant 
so that the latter may be considered to be small.

The decomposition of a general path into the classical path and fluctuations
around it as employed in (\ref{gli:eq:clfluc}) in the previous section was
merely a matter of convenience. For the exactly solvable case of a driven
harmonic oscillator it is not really relevant how we express a general path
satisfying the boundary conditions. Within the semiclassical approximation,
however, it is decisive to expand around the path leading to the dominant
contribution, i.e. the classical path. From a more mathematical point of
view, we have to evaluate a path integral over $\exp(\I S/\hbar)$ for
small $\hbar$. This can be done in a systematic way by the method of stationary
phase\index{stationary phase!method of} where the exponent has to be expanded 
around the extrema of the action $S$.

\begin{figure}[t]
\begin{center}
\includegraphics[width=0.6\textwidth]{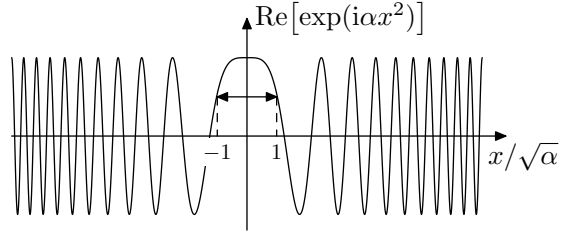}
\end{center}
\caption{In stationary phase approximation only a small region around the
extremum contributes to the integral. For the example shown here, the extremum
lies at $x=0$}
\label{gli:fig:statphapprox}
\end{figure}

At this point it may be useful to give a brief reminder of the method of 
stationary phase. Suppose we want to evaluate the integral
\begin{equation}
I(\alpha) = \int_{-\infty}^{\infty}\D x g(x)\exp\!\big(\I \alpha 
f(x)\big)
\end{equation}
in the limit of very large $\alpha$. Inspection of 
Fig.~\ref{gli:fig:statphapprox}, where $f(x)=x^2$, suggests that the dominant 
contribution to the integral comes from a region, in our example of size 
$1/\sqrt{\alpha}$, around the extremal (or stationary) point of the function 
$f(x)$. Outside of this region, the integrand is rapidly oscillating and 
therefore gives to leading order a negligible contribution. Since for large 
$\alpha$, the region determining the integral is very small, we may expand the 
function $f(x)$ locally around the extremum $x_0$
\begin{equation}
f(x)\approx f(x_0) + \frac{1}{2}f''(x_0)(x-x_0)^2+\dots
\end{equation}
and replace $g(x)$ by $g(x_0)$. Neglecting higher order terms, which is allowed 
if $f''(x_0)$ is of order one, we are left with the Gaussian integral
\begin{equation}
\begin{aligned}
I(\alpha)&\approx g(x_0)\exp\!\big(\I \alpha f(x_0)\big)
\int_{-\infty}^{\infty}\D x\exp\!\left(\frac\I {2}
f''(x_0)(x-x_0)^2\right)\\
&= \sqrt{\frac{2\pi}{\vert f''(x_0)\vert}}g(x_0)\exp\!\left[\I\alpha 
f(x_0)+\I\frac{\pi}{4}\mbox{sgn}\big(f''(x_0)\big)\right]\;,
\end{aligned}
\end{equation}
where $\mbox{sgn}(f''(x_0))$ denotes the sign of $f''(x_0)$. If $f(x)$ 
possesses more than one extremum, one has to sum over the contributions of all 
extrema unless one extremum can be shown to be dominant.

We now apply the stationary phase approximation to path integrals where
$1/\hbar$ plays the role of the large parameter. Since the action is stationary
at classical paths, we are obliged to express the general path as
\begin{equation}
x(s) = x_{\rm cl}(s)+\xi(s)\;,
\end{equation}
where $x_{\rm cl}$ is the classical path (or one of several possible paths)
satisfying the boundary conditions and $\xi$ represents the fluctuations
around the classical path. With this decomposition the action becomes
\begin{equation}
\begin{aligned}
S &= \int_0^t\D s\left(\frac{m}{2}\dot x^2-V(x)\right)\\
  &= \int_0^t\D s\left(\frac{m}{2}\dot x_{\rm cl}^2-V(x_{\rm cl})\right)
     +\int_0^t\D s\left(m\dot x_{\rm cl}\dot\xi-V'(x_{\rm cl})\xi\right)\\
&\qquad\qquad +\int_0^t\D s\left(\frac{m}{2}\dot\xi^2-
  \frac{1}{2}V''(x_{\rm cl})\xi^2\right)+\dots
\label{gli:eq:decomp}
\end{aligned}
\end{equation}
It is instructive to compare this result with the action (\ref{gli:eq:sdo})
for the driven harmonic oscillator. Again, the first term represents the
classical action. The second term vanishes as was shown explicitly in
(\ref{gli:eq:sdopi}) for the driven oscillator. In the general case, one
can convince oneself by partial integration of the kinetic part and comparison
with the classical equation of motion that this term vanishes again. This
is of course a consequence of the fact that the classical path, around which
we expand, corresponds to an extremum of the action. The third term on the
right-hand-side of (\ref{gli:eq:decomp}) is the leading order term in the
fluctuations as was the case in (\ref{gli:eq:sdo}). There is however an 
important difference since for anharmonic potentials the second derivative of 
the potential $V''$ is not constant and therefore the contribution of the
fluctuations depends on the classical path. Finally, in general there will
be higher order terms in the fluctuations as indicated by the dots
in (\ref{gli:eq:decomp}). The semiclassical approximation consists in 
neglecting these higher order terms so that after a partial integration, we get
for the action
\begin{equation}
S_{\rm sc} = S_{\rm cl}-\frac{1}{2}\int_0^t\D s\,\xi\!\left(m\frac{\D^2}
{\D s^2}+V''(x_{\rm cl})\right)\!\xi
\label{gli:eq:scaction}
\end{equation}
where the index `sc' indicates the semiclassical 
approximation.\index{semiclassical approximation}

Before deriving the propagator in semiclassical approximation, we have to
discuss the regime of validity of this approximation. Since the first term
in (\ref{gli:eq:scaction}) gives only rise to a global phase factor, it is
the second term which determines the magnitude of the quantum 
fluctuations\index{quantum fluctuations}.
For this term to contribute, we should have $\xi^2/\hbar\lesssim1$ so that
the magnitude of typical fluctuations is at most of order $\sqrt{\hbar}$. The 
term of third order in the fluctuations is already smaller than the second 
order term by a factor $(\sqrt{\hbar})^3/\hbar = \sqrt{\hbar}$. If Planck's 
constant can be considered to be small, we may indeed neglect the fluctuation 
contributions of higher than second order except for one exception: It may 
happen that the second order term does not contribute, as has been the case at 
the conjugate points for the driven harmonic oscillator in 
Sect.~\ref{gli:subsec:dho}. Then, the leading nonvanishing contribution becomes 
dominant. For the following discussion, we will not consider this latter case.

In analogy to Sect.~\ref{gli:subsec:dho} we obtain for the propagator in
semiclassical approximation
\begin{equation}
K(x_{\rm f},t,x_{\rm i} ,0) = \sqrt{\frac{m}{2\pi\I\hbar t}}
\sqrt{\frac{D_0}{D}}\exp\!\left(\frac\I {\hbar}S_{\rm cl}\right)
\end{equation}
where 
\begin{equation}
D=\det\left(\frac{\D^2}{\D s^2}+V''(x_{\rm cl})\right)
\end{equation}
and $D_0$ is the fluctuation determinant (\ref{gli:eq:det0}) of the free 
particle.

Even though it may seem that determining the prefactor is a formidable task
since the fluctuation determinant for a given potential has to be evaluated,
this task can be greatly simplified. In addition, the following considerations
offer the benefit of providing a physical interpretation of the prefactor. In 
our evaluation of the prefactor we follow Marinov \cite{gli:marin80}. The main 
idea is to make use of the semigroup property (\ref{gli:eq:semigroup}) of the 
propagator 
\begin{align}
\label{gli:eq:scsemigroup}
&C(x_{\rm f},t,x_{\rm i},0)\exp\!\left[\frac{\rm i}{\hbar}S_{\rm cl}
(x_{\rm f},t,x_{\rm i},0)\right]\\
&= \int\D x'
C(x_{\rm f},t,x',t')C(x',t',x_{\rm i},0)\exp\!\left[\frac{\rm i}{\hbar}
\big[S_{\rm cl}(x_{\rm f},t,x',t')+ S_{\rm cl}(x',t',x_{\rm i},0)\big]\right]
\nonumber
\end{align}
where the prefactor $C$ depends on the fluctuation contribution. We now have to 
evaluate the $x'$-integral within the semiclassical approximation. According to 
the stationary phase requirement discussed above, the dominant contribution to 
the integral comes from $x'=x_0(x_{\rm f}, x_{\rm i}, t, t')$ satisfying
\begin{equation}
\left.\frac{\partial S_{\rm cl}(x_{\rm f},t,x',t')}{\partial x'}
\right\vert_{x'=x_0} + \left.\frac{\partial S_{\rm cl}(x',t',x_{\rm i},0)}
{\partial x'}\right\vert_{x'=x_0} =0\;.
\label{gli:eq:statphase}
\end{equation}
According to classical mechanics these derivatives are related to initial
and final momentum by \cite{gli:brack97}
\begin{equation}
\left(\frac{\partial S_{\rm cl}}{\partial x_{\rm i}}\right)_{x_{\rm f},
t_{\rm f}, t_{\rm i}} = -p_{\rm i}\qquad\qquad
\left(\frac{\partial S_{\rm cl}}{\partial x_{\rm f}}\right)_{x_{\rm i},
t_{\rm f}, t_{\rm i}} = p_{\rm f}
\end{equation}
so that (\ref{gli:eq:statphase}) can expressed as
\begin{equation}
p(t'-\varepsilon) = p(t'+\varepsilon)\;.
\end{equation}
The point $x_0$ thus has to be chosen such that the two partial classical
paths can be joined with a continuous momentum. Together they therefore yield
the complete classical path and in particular
\begin{equation}
S_{\rm cl}(x_{\rm f},t,x_{\rm i},0) = S_{\rm cl}(x_{\rm f},t,x_0,t') +
S_{\rm cl}(x_0,t',x_{\rm i},0)\;.
\label{gli:eq:sumaction}
\end{equation}
This relation ensures that the phase factors depending on the classical actions 
on both sides of (\ref{gli:eq:scsemigroup}) are equal.

After having identified the stationary path, we have to evaluate the integral 
over $x'$ in (\ref{gli:eq:scsemigroup}). Within semiclassical approximation
this Gaussian integral leads to
\begin{align}
\label{gli:eq:preact}
&\frac{C(x_{\rm f},t,x_{\rm i},0)}{C(x_{\rm f},t,x_0,t')C(x_0,t',x_{\rm i},0)}\\
&\hspace{3truecm}= \left(\frac{1}{2\pi\I \hbar}\frac{\partial^2}
{\partial x_0^2}\big[S_{\rm cl}(x_{\rm f},t,x_0,t')+
S_{\rm cl}(x_0,t',x_{\rm i},0)\big]\right)^{-1/2}.\nonumber
\end{align}
In order to make progress, it is useful to take the derivative of 
(\ref{gli:eq:sumaction}) with respect to $x_{\rm f}$ and $x_{\rm i}$. Keeping
in mind that $x_0$ depends on these two variables one finds
\begin{equation}
\begin{aligned}
\frac{\partial^2 S_{\rm cl}(x_{\rm f},t,x_{\rm i},0)}
{\partial x_{\rm f}\partial x_{\rm i}} &=
\frac{\partial^2 S_{\rm cl}(x_{\rm f},t,x_0,t')}
{\partial x_{\rm f}\partial x_0}\frac{\partial x_0}{\partial x_{\rm i}} +
\frac{\partial^2 S_{\rm cl}(x_0,t',x_{\rm i},0)}
{\partial x_{\rm i}\partial x_0} \frac{\partial x_0}{\partial x_{\rm f}}\\
&\qquad+ \frac{\partial^2}{\partial x_0^2}\big[S_{\rm cl}(x_{\rm f},t,x_0,t') +
S_{\rm cl}(x_0,t',x_{\rm i},t)\big]\frac{\partial x_0}{\partial x_{\rm i}}
\frac{\partial x_0}{\partial x_{\rm f}}\;.
\end{aligned}
\label{gli:eq:partderi}
\end{equation}
Similarly, one finds by taking derivatives of the stationary phase condition 
(\ref{gli:eq:statphase}) 
\begin{equation}
\frac{\partial x_0}{\partial x_{\rm f}} = 
-\frac{\dfrac{\partial^2}{\partial x_{\rm f}x_0}S_{\rm cl}(x_{\rm f},t,x_0,t')}
{\dfrac{\partial^2}{\partial x_0^2}\big[S_{\rm cl}(x_{\rm f},t,x_0,t')
+S_{\rm cl}(x_0,t',x_{\rm i},0)\big]}
\end{equation}
and
\begin{equation}
\frac{\partial x_0}{\partial x_{\rm i}} = 
-\frac{\dfrac{\partial^2}{\partial x_{\rm i}x_0}S_{\rm cl}(x_0,t',x_{\rm i},0)}
{\dfrac{\partial^2}{\partial x_0^2}\big[S_{\rm cl}(x_{\rm f},t,x_0,t')
+S_{\rm cl}(x_0,t',x_{\rm i},0)\big]}\;.
\end{equation}
These expressions allow to eliminate the partial derivatives of $x_0$ with 
respect to $x_{\rm i}$ and $x_{\rm f}$ appearing in (\ref{gli:eq:partderi}) and 
one finally obtains
\begin{align}
&\left(\dfrac{\partial^2}{\partial x_0^2}\big[ 
S_{\rm cl}(x_{\rm f},t,x_0,t')+S_{\rm cl}(x_0,t',x_{\rm i},0)\big]\right)^{-1}\\
&\hspace{3truecm}=-\frac{\dfrac{\partial^2}{\partial x_{\rm i}
\partial x_{\rm f}}S_{\rm cl}(x_{\rm f},t,x_{\rm i},0)}
{\dfrac{\partial^2 S_{\rm cl}(x_{\rm f},t,x_0,t')}
{\partial x_{\rm f}\partial x_0}\dfrac{\partial^2 
S_{\rm cl}(x_0,t',x_{\rm i},0)}{\partial x_{\rm i}\partial x_0}}\nonumber\;.
\end{align}
Inserting this result into (\ref{gli:eq:preact}), the prefactor can be 
identified as the so-called Van Vleck-Pauli-Morette 
determinant\index{Van Vleck-Pauli-Morette determinant}\index{determinant!Van 
Vleck-Pauli-Morette} \cite{gli:vleck28,gli:moret51,gli:pauli00}
\begin{equation}
C(x_{\rm f},t,x_{\rm i},0) = \left[\frac{1}{2\pi{\rm i}\hbar}\left(
-\frac{\partial^2 S_{\rm cl}(x_{\rm f},t,x_{\rm i},0)}{\partial x_{\rm f}
\partial x_{\rm i}}\right)\right]^{1/2}
\end{equation}
so that the propagator in semiclassical approximation finally 
reads\index{propagator!in semiclassical approximation}\index{semiclassical 
approximation}
\begin{align}
\label{gli:eq:scprop}
&K(x_{\rm f},t,x_{\rm i},0)\\ 
&\hspace{1truecm}= \left(\frac{1}{2\pi\hbar}\left\vert
-\frac{\partial^2 S_{\rm cl}(x_{\rm f},t,x_{\rm i},0)}{\partial x_{\rm f}
\partial x_{\rm i}}\right\vert\right)^{1/2}\exp\!\left[\frac{\rm i}{\hbar} 
S_{\rm cl}(x_{\rm f},t,x_{\rm i},0)-{\rm i}\left(\frac{\pi}{4}+
n\frac{\pi}{2}\right)\right]\nonumber
\end{align}
where the Morse index\index{Morse index} $n$ denotes the number of sign 
changes of $\partial^2 S_{\rm cl}/\partial x_{\rm f}\partial x_{\rm i}$
\cite{gli:morse66}. We had encountered such a phase factor before in the 
propagator (\ref{gli:eq:propdho}) of the harmonic oscillator.

As we have already mentioned above, derivatives of the action with respect
to position are related to momenta. This allows to give a physical 
interpretation of the prefactor of the propagator as the change of the end 
point of the path as a function of the initial momentum
\begin{equation}
\left(-\frac{\partial^2 S_{\rm cl}}{\partial x_{\rm i}\partial x_{\rm
f}}\right)^{-1} =\frac{\partial x_{\rm f}}{\partial p_{\rm i}}\;.
\end{equation}
A zero of this expression, or equivalently a divergence of the prefactor of the
propagator, indicates a conjugate point\index{conjugate point} where the end 
point does not depend on the initial momentum.

To close this section, we want to compare the semiclassical result 
(\ref{gli:eq:scprop}) with exact results for the free particle and the
harmonic oscillator. In our discussion of the free particle in 
Sect.~\ref{gli:subsec:freeparticle} we already mentioned that the propagator 
can be expressed entirely in terms of classical quantities. Indeed, the
expression (\ref{gli:eq:kfpcl}) for the propagator of the free particle agrees
with (\ref{gli:eq:scprop}).

For the harmonic oscillator, we know from Sect.~\ref{gli:subsec:dho} that the
prefactor does not depend on a possibly present external force. We may 
therefore consider the action (\ref{gli:eq:dhoaction}) in the absence of 
driving $f(s)=0$ which then reads
\begin{equation}
S_{\rm cl}=\frac{m\omega}{2\sin(\omega t)}\left[\left(x_{\rm i}^2+x_{\rm f}^2
\right)\cos(\omega t)-2x_{\rm i}x_{\rm f}\right]\;.
\end{equation}
Taking the derivative with respect to $x_{\rm i}$ and $x_{\rm f}$ one finds
for the prefactor
\begin{equation}
C(x_{\rm f},t,x_{\rm i},0) = \left(\frac{m\omega}{2\pi{\rm i}\hbar
\sin(\omega t)}\right)^{1/2}
\end{equation}
which is identical with the prefactor in our previous result 
(\ref{gli:eq:propdho}). As expected, for potentials at most quadratic in the 
coordinate the semiclassical propagator agrees with the exact expression.

\subsection{Imaginary Time Path Integral}
\label{gli:subsec:itpi}

In the discussion of dissipative systems we will be dealing with a system
coupled to a large number of environmental degrees of freedom. In most cases,
the environment will act like a large heat bath characterized by a temperature 
$T$. The state of the environment will therefore be given by an equilibrium
density matrix. Occasionally, we may also be interested in the equilibrium
density matrix of the system itself. Such a state may be reached after 
equilibration due to weak coupling with a heat bath. 

In order to describe such thermal equilibrium states and the dynamics of the 
system on a unique footing, it is desirable to express equilibrium density 
matrices in terms of path integrals. This is indeed possible as one recognizes
by writing the equilibrium density operator in position 
representation\index{equilibrium density matrix}
\begin{equation}
\rho_{\beta}(x,x') = \frac{1}{\mathcal Z}\langle x\vert \exp(-\beta H)\vert
x'\rangle
\end{equation}
with the partition function\index{partition function}
\begin{equation}
\mathcal{Z}=\int\!\D x\langle x\vert\exp(-\beta H)\vert x\rangle\;.
\end{equation}
Comparing with the propagator in position representation
\begin{equation}
K(x,t,x',0) = \langle x\vert\exp\!\left(-\frac\I {\hbar}Ht\right)\vert x'
\rangle
\end{equation}
one concludes that apart from the partition function the equilibrium density 
matrix is equivalent to a propagator in imaginary time $t=-\I \hbar\beta$.

After the substitution $\sigma=\I s$ the action in imaginary time 
$-\I \hbar\beta$ reads
\begin{equation}
 \int_0^{-\I \hbar\beta}\D s\left[\frac{m}{2}\left(\frac{\D x}
{\D s}\right)^2-V(x)\right]
= \I \int_0^{\hbar\beta}\D \sigma\left[\frac{m}{2}
\left(\frac{\D x}{\D\sigma}\right)^2+V(x)\right]\;.
\label{gli:eq:imagtime}
\end{equation}
Here and in the following, we use greek letters to indicate imaginary 
times. Motivated by the right-hand side of (\ref{gli:eq:imagtime}) we define
the so-called Euclidean action\index{Euclidean action}\index{action!Euclidean}
\begin{equation}
S^{\rm E}[x] = \int_0^{\hbar\beta}\D \sigma\left[\frac{m}{2}\dot x^2+V(x)
\right]\;.
\label{gli:eq:euclidean}
\end{equation}
Even though one might fear a lack of intuition for motion in imaginary time,
this results shows that it can simply be thought of as motion in the inverted 
potential in real time. With the Euclidean action (\ref{gli:eq:euclidean}) we 
now obtain as an important result the path integral expression for the 
(unnormalized) equilibrium density matrix\index{path integral representation!of
equilibrium density matrix}
\begin{equation}
\langle x\vert\exp(-\beta H)\vert x'\rangle = 
\int_{\bar x(0)=x'}^{\bar x(\hbar\beta)=x}{\cal D}\bar x 
\exp\!\left(-\frac{1}{\hbar} S^{\rm E}[\bar x]\right)\;.
\end{equation}
This kind of functional integral was discussed as early as 1923 by Wiener 
\cite{gli:wiene23} in the context of classical Brownian motion.

As an example we consider the (undriven) harmonic oscillator. There is actually
no need to evaluate a path integral since we know already from 
Sect.~\ref{gli:subsec:dho} the propagator
\begin{equation}
K(x_{\rm f},t,x_{\rm i},0) = \sqrt{\frac{m\omega}{2\pi{\rm i}\hbar 
\sin(\omega t)}} \exp\!\left[-\I \frac{m\omega}{2\hbar}
\frac{(x_{\rm i}^2 + x_{\rm f}^2)\cos(\omega t) - 2x_{\rm i}x_{\rm f}}
{\sin(\omega t)}\right]\;.
\end{equation}
Transforming the propagator into imaginary time $t\to -\I \hbar\beta$ and
renaming $x_{\rm i}$ and $x_{\rm f}$ into $x'$ and $x$, respectively, one 
obtains the equilibrium density matrix
\begin{align}
\label{gli:eq:dmho}
&\rho_{\beta}(x ,x')\\
&\qquad= \frac{1}{\mathcal Z}
\sqrt{\frac{m\omega}{2\pi\hbar\sinh(\hbar\beta\omega)}}
\exp\!\left[-\frac{m\omega}{2\hbar}\frac{(x^2 + x'^2)\cosh(\hbar\beta\omega) - 
2xx'}{\sinh(\hbar\beta\omega)}\right]\nonumber\;.
\end{align}
The partition function is obtained by performing the trace as
\begin{equation}
\mathcal{Z} = \int\D x \langle x\vert\exp(-\beta H)\vert x\rangle=
\frac{1}{2\sinh(\hbar\beta\omega/2)}
\label{gli:eq:partho}
\end{equation}
which agrees with the expression
\begin{equation}
\mathcal{Z}=\sum_{n=0}^{\infty}\exp\!\left[-\beta\hbar\omega\left(n+
\dfrac{1}{2}\right)\right]
\end{equation}
based on the energy levels of the harmonic oscillator.

Since the partition function\index{partition function} often serves as a 
starting point for the calculation of thermodynamic properties, it is 
instructive to take a closer at how this quantity may be obtained within the 
path integral formalism. A possible approach is the one we just have sketched. 
By means of an imaginary time path integral one first calculates 
$\langle x\vert\exp(-\beta H)\vert x\rangle$ which is proportional to the 
probability to find the system at position $x$. Subsequent integration over 
coordinate space then yields the partition function. 

However, the partition function may also be determined in one step. To this
end, we expand around the periodic trajectory with extremal Euclidean action
which in our case is given by $x(\sigma)=0$. Any deviation will increase both 
the kinetic and potential energy and thus increase the Euclidean action. All 
other trajectories contributing to the partition function are generated by a 
Fourier series on the imaginary time interval from $0$ to $\hbar\beta$ 
\label{gli:page:zfou}
\begin{equation}
x(\sigma) = \frac{1}{\sqrt{\hbar\beta}}\left[a_0 + \sqrt{2}\sum_{n=1}^{\infty
}\big(a_n \cos(\nu_n\sigma) + b_n\sin(\nu_n\sigma)\big)\right]
\label{gli:eq:partans}
\end{equation}
where we have introduced the so-called Matsubara frequencies\index{Matsubara
frequency}
\begin{equation}
\nu_n = \frac{2\pi}{\hbar\beta}n\;.
\label{gli:eq:matsubara}
\end{equation}
This ansatz should be compared with (\ref{gli:eq:fourier}) for the 
fluctuations where $a_0$ was fixed because the fluctuations had to vanish
at the boundaries. For the partition function this requirement is dropped since 
we have to integrate over all periodic trajectories. Furthermore, we note that
indeed with the ansatz (\ref{gli:eq:partans}) only the periodic trajectories 
contribute.  All other paths cost an infinite amount of action due to the jump 
at the boundary as we will see shortly.

Inserting the Fourier expansion (\ref{gli:eq:partans}) into the Euclidean 
action of the harmonic oscillator
\begin{equation}
S^{\rm E} = \int_0^{\hbar\beta}\D \sigma\frac{m}{2}\left(\dot x^2 +
\omega^2x^2\right)
\end{equation}
we find
\begin{equation}
S^{\rm E} = \frac{m}{2}\!\left[\omega^2a_0^2 + \sum_{n=1}^{\infty}(\nu_n^2 +
\omega^2)(a_n^2+b_n^2)\right]\;.
\end{equation}
As in Sect.~\ref{gli:subsec:dho} we do not want to go into the mathematical 
details of integration measures and Jacobi determinants. Unfortunately, the
free particle cannot serve as a reference here because its partition function
does not exist. We therefore content ourselves with remarking that because of
\begin{equation}
\frac{1}{\omega}\prod_{n=1}^{\infty}\frac{1}{\nu_n^2+\omega^2} =
\frac{\hbar\beta}{\sum_{n=1}^{\infty}\nu_n^2}
\frac{1}{2\sinh(\hbar\beta\omega/2)}
\label{gli:eq:sinhrel}
\end{equation}
the result of the Gaussian integral over the Fourier coefficients yields the 
partition function up to a frequency independent factor. This enables us to
determine the partition function in more complicated cases by proceeding as
above and using the partition function harmonic oscillator as a reference.

Returning to the density matrix of the harmonic oscillator we finally obtain by
inserting the partition function (\ref{gli:eq:partho}) into the expression
(\ref{gli:eq:dmho}) for the density matrix\index{equilibrium density matrix!of
harmonic oscillator}
\begin{align}
\label{gli:eq:dmhof}
\rho_{\beta}(x,x') =& 
\sqrt{\frac{m\omega}{\pi\hbar}\frac{\cosh(\hbar\beta\omega)-1}
{\sinh(\hbar\beta\omega)}}\\
&\qquad\times\exp\!\left[-\frac{m\omega}{2\hbar}\frac{(x^2 + x'^2)
\cosh(\hbar\beta\omega) - 2xx'}{\sinh(\hbar\beta\omega)}\right]\;.\nonumber
\end{align}
Without path integrals, this result would require the evaluation of sums over 
Hermite polynomials.

The expression for the density matrix (\ref{gli:eq:dmhof}) can be verified in 
the limits of high and zero temperature. In the classical limit of very high 
temperatures, the probability distribution in real space is given by
\begin{equation}
P(x) = \rho_{\beta}(x,x) = \sqrt{\frac{\beta m\omega^2}{2\pi}}
\exp\!\left(-\beta\frac{m\omega^2}{2}x^2\right)\sim\exp[-\beta V(x)]\;.
\end{equation}
We thus have obtained the Boltzmann distribution which depends only on the
potential energy. The fact that the kinetic energy does not play a role can 
easily be understood in terms of the path integral formalism. Excursions in a
very short time $\hbar\beta$ cost too much action and are therefore strongly
suppressed. 

In the opposite limit of zero temperature the density matrix factorizes into
a product of ground state wave functions of the harmonic oscillator
\begin{equation}
\lim_{\beta\to\infty}\rho_{\beta}(x,x') =
\left[\left(\frac{m\omega}{\pi\hbar}\right)^{1/4}\exp\!\left(-\frac{m\omega}
{2\hbar}x^2\right)\right]
\left[\left(\frac{m\omega}{\pi\hbar}\right)^{1/4}\exp\!\left(-\frac{m\omega}
{2\hbar}x'^2\right)\right]
\end{equation}
as should be expected.

\section{Dissipative Systems}
\label{gli:sec:dissipation}

\subsection{Introduction}
In classical mechanics dissipation can often be adequately described by
including a velocity dependent damping term into the equation of motion. Such 
a phenomenological approach is no longer possible in quantum mechanics where
the Hamilton formalism implies energy conservation for time-independent 
Hamiltonians. Then, a better understanding of the situation is necessary in
order to arrive at an appropriate physical model. 

A damped pendulum may help us to understand the mechanism of dissipation.
The degree of freedom of interest, the elongation of the pendulum, undergoes
a damped motion because it interacts with other degrees of freedom, the 
molecules in the air surrounding the pendulum's mass. We may consider the
pendulum and the air molecules as one large system which, if assumed to be
isolated from further degrees of freedom, obeys energy conservation. The energy
of the pendulum alone, however, will in general not be conserved. This single
degree of freedom is therefore subject to dissipation arising from the coupling
to other degrees of freedom.

This insight will allow us in the following section to introduce a model
for a system coupled to an environment and to demonstrate explicitly its
dissipative nature. In particular, we will introduce the quantities needed 
for a description which focuses on the system degree of freedom. We are then
in a position to return to the path integral formalism and to demonstrate how
it may be employed to study dissipative systems. Starting from the model of 
system and environment, the latter will be eliminated to obtain a reduced 
description for the system alone. This leaves us with an effective action which 
forms the basis of the path integral description of dissipation.

\subsection{Environment as Collection of Harmonic Oscillators}
\label{gli:sec:model}

A suitable model for dissipative quantum systems should both incorporate
the idea of a coupling between system and environment and be amenable to
an analytic treatment of the environmental coupling. These requirements are met 
by a model which nowadays is often referred to as Caldeira-Leggett 
model\index{Caldeira-Leggett model} \cite{gli:calde81,gli:calde83} even though 
it has been discussed in the literature under various names before for harmonic 
systems \cite{gli:magal59,gli:senit60,gli:ford65,gli:uller66} and anharmonic 
systems \cite{gli:zwanz73}. The Hamiltonian 
\begin{equation}
H=H_{\rm S} + H_{\rm B} + H_{\rm SB}
\label{gli:eq:clham}
\end{equation}
consists of three contributions. The Hamiltonian of the system degree of 
freedom 
\begin{equation}
H_{\rm S} = \frac{p^2}{2m} + V(q)
\end{equation}
models a particle of mass $m$ moving in a potential $V$. Here, we denote the
coordinate by $q$ to facilitate the distinction from the environmental 
coordinates $x_n$ which we will introduce in a moment. Of course, the system 
degree of freedom does not have to be associated with a real particle but may 
be quite abstract. In fact, a substantial part of the calculations to be 
discussed in the following does not depend on the detailed form of the system 
Hamiltonian.

The Hamiltonian of the environmental degrees of freedom
\begin{equation}
H_{\rm B} = \sum_{n=1}^N\left(\frac{p_n^2}{2m_n} + 
\frac{m_n}{2}\omega_n^2x_n^2\right)
\label{gli:eq:hb}
\end{equation}
describes a collection of harmonic oscillators. While the properties of the 
environment may in some cases be chosen on the basis of a microscopic model, 
this does not have to be the case. Often, a phenomenological approach is 
sufficient as we will see below. As an example we mention an Ohmic resistor 
which as a linear electric element should be well described by a Hamiltonian of 
the form (\ref{gli:eq:hb}). On the other hand, the underlying mechanism leading 
to dissipation, e.g. in a resistor, may be much more complicated than that
implied by the model of a collection of harmonic oscillators.

The coupling defined by the Hamiltonian
\begin{equation}
H_{\rm SB} = -q\sum_{n=1}^N c_nx_n + q^2\sum_{n=1}^N
\frac{c_n^2}{2m_n\omega_n^2}
\label{gli:eq:hsb}
\end{equation}
is bilinear in the position operators of system and environment.
There are cases where the bilinear coupling is realistic, e.g.\ for an 
environment consisting of a linear electric circuit like the resistor just
mentioned or for a dipolar coupling to electromagnetic field modes encountered 
in quantum optics. Within a more general scope, this Hamiltonian may be viewed 
as linearization of a nonlinear coupling in the limit of weak coupling to the 
environmental degrees of freedom. As was first pointed out by Caldeira and 
Leggett, an infinite number of degrees of freedom still allows for strong 
damping even if each environmental oscillator couples only weakly to the system 
\cite{gli:calde81,gli:calde83}.

An environment consisting of harmonic oscillators as in (\ref{gli:eq:hb}) might
be criticized. If the potential $V(q)$ is harmonic, one may pass to normal
coordinates and thus demonstrate that after some time a revival of the initial
state will occur. For sufficiently many environmental oscillators, however,
this so-called Poincar\'e recurrence time\index{Poincar{\'e} recurrence time} 
tends to infinity \cite{gli:hemme58}. Therefore, even with a linear environment 
irreversibility becomes possible at least for all practical purposes.

The reader may have noticed that in the coupling Hamiltonian (\ref{gli:eq:hsb})
a term is present which only contains an operator acting in the system Hilbert 
space but depends on the coupling constants $c_n$. The physical reason for the
inclusion of this term lies in a potential renormalization\index{potential
renormalization} introduced by the first term in (\ref{gli:eq:hsb}). This 
becomes clear if we consider the minimum of the Hamiltonian with respect to the 
system and environment coordinates. From the requirement
\begin{equation}
\frac{\partial H}{\partial x_n} = m_n\omega_n^2x_n - c_nq \stackrel{!}{=} 0
\end{equation}
we obtain
\begin{equation}
x_n = \frac{c_n}{m_n\omega_n^2}q\;.
\end{equation}
Using this result to determine the minimum of the Hamiltonian with respect
to the system coordinate we find
\begin{equation}
\frac{\partial H}{\partial q} = \frac{\partial V}{\partial q} -
\sum_{n=1}^N c_nx_n + q\sum_{n=1}^N \frac{c_n^2}{m_n\omega_n^2} =
\frac{\partial V}{\partial q}\;. 
\end{equation}
The second term in (\ref{gli:eq:hsb}) thus ensures that this minimum is 
determined by the bare potential $V(q)$. 

After having specified the model, we now want to derive an effective description
of the system alone. It was first shown by Magalinski{\u\i} \cite{gli:magal59} 
that the elimination\index{elimination of environment} of the environmental 
degrees of freedom leads indeed to a damped equation of motion for the system 
coordinate. We perform the elimination within the Heisenberg picture where the 
evolution of an operator $A$ is determined by
\begin{equation}
\frac{\D A}{\D t} = \frac\I {\hbar}[H,A]\;.
\end{equation}
From the Hamiltonian (\ref{gli:eq:clham}) we obtain the equations of motion
for the environmental degrees of freedom
\begin{equation}
\begin{aligned}
\dot p_n &= -m_n\omega_n^2 x_n + c_n q\\
\dot x_n &= \frac{p_n}{m_n}
\label{gli:eq:emenv}
\end{aligned}
\end{equation}
and the system degree of freedom
\begin{equation}
\begin{aligned}
\dot p &= -\frac{\partial V}{\partial q} +\sum_{n=1}^N c_n x_n 
-q\sum_{n=1}^N\frac{c_n^2}{m_n\omega_n^2}\\
\dot x &= \frac{p}{m}\;.
\label{gli:eq:emsys}
\end{aligned}
\end{equation}

The trick for solving the environmental equations of motion (\ref{gli:eq:emenv})
consists in treating the system coordinate $q(t)$ as if it were a given 
function of time. The inhomogeneous differential equation then has the solution
\begin{equation}
x_n(t)=x_n(0)\cos(\omega_n t) + \frac{p_n(0)}{m_n\omega_n}\sin(\omega_n t)
+\frac{c_n}{m_n\omega_n}\int_0^t\D s\sin\big(\omega_n(t-s)\big)q(s)\;.
\end{equation}
Inserting this result into (\ref{gli:eq:emsys}) one finds an effective equation
of motion for the system coordinate
\begin{align}
&m\ddot q - \int_0^t\D s\sum_{n=1}^N\frac{c_n^2}{m_n\omega_n}
\sin\big(\omega_n(t-s)\big)q(s) + \frac{\partial V}{\partial q} 
+q\sum_{n=1}^N\frac{c_n^2}{m_n\omega_n^2}\\
&\hspace{5truecm}=\sum_{n=1}^Nc_n\left[x_n(0)\cos(\omega_n t)+
\frac{p_n(0)}{m_n\omega_n} \sin(\omega_n t)\right].\nonumber
\end{align}
By a partial integration of the second term on the left-hand side this 
equation of motion can be cast into its final form\index{effective equation of
motion}
\begin{equation}
m\ddot q + m\int_0^t\D s\gamma(t-s)\dot q(s) + \frac{\partial V}{\partial
q} = \xi(t)
\label{gli:eq:effem}
\end{equation}
with the damping kernel\index{damping kernel}
\begin{equation}
\gamma(t)=\frac{1}{m}\sum_{n=1}^N\frac{c_n^2}{m_n\omega_n^2}\cos(\omega_n t)
\end{equation}
and the operator-valued fluctuating force\index{fluctuating force}
\begin{equation}
\xi(t) =
\sum_{n=1}^Nc_n\left[\left(x_n(0)-\frac{c_n}{m_n\omega_n^2}q(0)\right)
\cos(\omega_n t) + \frac{p_n(0)}{m_n\omega_n}\sin(\omega_n t)\right]\;.
\label{gli:eq:noise}
\end{equation}

The fluctuating force vanishes if averaged over a thermal density matrix of
the environment including the coupling to the system
\begin{equation}
\langle\xi(t)\rangle_{\rm B+SB}=\frac{\mathrm{Tr}_{\rm B}\big[\xi(t)
\exp\!\big({-}\beta(H_{\rm B}+H_{\rm SB})\big)\big]}{\mathrm{Tr}_{\rm B}
\big[\exp\!\big({-}\beta(H_{\rm B}+H_{\rm SB})\big)\big]} = 0\;.
\end{equation}
For weak coupling, one may want to split off the transient term 
$m\gamma(t)q(0)$ which is of second order in the coupling and write the 
fluctuating force as \cite{gli:hangg97}
\begin{equation}
\xi(t)=\zeta(t)-m\gamma(t)q(0)\;.
\label{gli:eq:zeta}
\end{equation}
The so defined force $\zeta(t)$ vanishes if averaged over the environment alone
\begin{equation}
\langle\zeta(t)\rangle_{\rm B}=\frac{\mathrm{Tr}_{\rm B}\big[\zeta(t)
\exp(-\beta H_{\rm B})\big]}{\mathrm{Tr}_{\rm B}\big[\exp(-\beta 
H_{\rm B})\big]} = 0\;.
\end{equation}

An important quantity to characterize the fluctuating force is the correlation
function which again can be evaluated for $\xi$ with respect to $H_{\rm B} +
H_{\rm SB}$ or equivalently for $\zeta$ with respect to $H_{\rm B}$ alone. With
(\ref{gli:eq:noise}) and (\ref{gli:eq:zeta}) we get the correlation function
\begin{equation}
\langle \zeta(t)\zeta(0)\rangle_{\rm B} = \sum_{n,l}c_nc_l\left\langle \left(
x_n(0)\cos(\omega_n t) + \frac{p_n(0)}{m_n\omega_i}\sin(\omega_n t)
\right)x_l(0)\right\rangle_{\rm B}\;.
\label{gli:eq:zetazeta}
\end{equation}
In thermal equilibrium the second moments are given by
\begin{align}
\left\langle x_n(0)x_l(0)\right\rangle_{\rm B} &=
\delta_{nl}\frac{\hbar}{2m_n\omega_n}\coth\left(\frac{\hbar\beta\omega_n}{2}
\right)\\[0.1truecm]
\left\langle p_n(0) x_l(0)\right\rangle_{\rm B} &=
-\frac{\I \hbar}{2}\delta_{nl}\;,
\end{align}
so that the noise correlation function\index{noise correlation function}
finally becomes
\begin{equation}
\langle\zeta(t)\zeta(0)\rangle_{\rm B} = \sum_{n=1}^N \frac{\hbar c_n^2}{2m_n
\omega_n}\left[\coth\left(\frac{\hbar\beta\omega_n}{2}\right)\cos(\omega_n t)
-\I \sin(\omega_n t)\right]\;.
\label{gli:eq:noisecorr}
\end{equation}
The imaginary part appearing here is a consequence of the fact that the
operators $\zeta(t)$ and $\zeta(0)$ in general do not commute. The correlation 
function (\ref{gli:eq:noisecorr})  appears as integral kernel both in master 
equations as well as in the effective action derived below (cf.\ 
(\ref{gli:eq:knkernel}) and (\ref{gli:eq:kntau})).

It is remarkable that within a reduced description for the system alone all
quantities characterizing the environment may be expressed in terms of the
spectral density of bath oscillators\index{spectral density of bath oscillators}
\begin{equation}
J(\omega) = \pi\sum_{n=1}^N \frac{c_n^2}{2m_n\omega_n}
\delta(\omega-\omega_n)\;.
\label{gli:eq:j}
\end{equation}
As an example, the damping kernel may be expressed in terms of this spectral
density as
\begin{equation}
\gamma(t)=\frac{1}{m}\sum_{n=1}^N\frac{c_n^2}{m_n\omega_n^2}\cos(\omega_n t)
= \frac{2}{m}\int_0^{\infty}\frac{\D \omega}{\pi}\frac{J(\omega)}
{\omega}\cos(\omega t)\;.
\label{gli:eq:gammaj}
\end{equation}
For practical calculations, it is therefore unnecessary to specify all
parameters $m_n, \omega_n$ and $c_n$ appearing in (\ref{gli:eq:hb}) and
(\ref{gli:eq:hsb}). It rather suffices to define the spectral density 
$J(\omega)$.

The most frequently used spectral density\index{spectral density of bath 
oscillators!for Ohmic damping}\index{Ohmic damping!spectral density of bath
oscillators for}
\begin{equation}
J(\omega) = m\gamma\omega
\label{gli:eq:johmic}
\end{equation}
is associated with the so-called Ohmic damping. This term is sometimes employed 
to indicate a proportionality to frequency merely at low frequencies instead of
over the whole frequency range. In fact, in any realistic situation the 
spectral density will not increase like in (\ref{gli:eq:johmic}) for 
arbitrarily high frequencies. It is justified to use the term ``Ohmic damping'' 
even if (\ref{gli:eq:johmic}) holds only below a certain frequency provided 
this frequency is much higher than the typical frequencies appearing in the system dynamics.

From (\ref{gli:eq:gammaj}) one finds the damping kernel for Ohmic 
damping\index{damping kernel!for Ohmic damping}\index{Ohmic damping!damping 
kernel for}
\begin{equation}
\gamma(t)=2\gamma\delta(t)\;,
\label{gli:eq:gammaohmic}
\end{equation}
which renders (\ref{gli:eq:effem}) memory-free. We thus recover the velocity 
proportional damping term familiar from classical damped systems. It should be 
noted that the factor of two in (\ref{gli:eq:gammaohmic}) disappears upon 
integration in (\ref{gli:eq:effem}) since (\ref{gli:eq:gammaj}) implies that 
the delta function is symmetric around zero.

\begin{figure}[t]
\begin{center}
\includegraphics[width=0.3\textwidth]{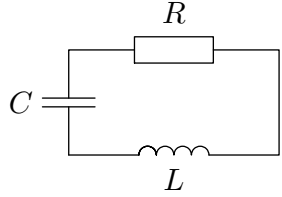}
\end{center}
\caption{$LC$ oscillator with Ohmic damping due to a resistor $R$}
\label{gli:fig:circuit}
\end{figure}

At this point, we want to briefly elucidate the origin of the term ``Ohmic 
damping''\index{Ohmic damping}. Let us consider the electric circuit shown in 
Fig.~\ref{gli:fig:circuit} consisting of a resistance $R$, a capacitance $C$ 
and an inductance $L$. Summing up the voltages around the loop, one obtains as 
equation of motion for the charge $Q$ on the capacitor
\begin{equation}
L\ddot Q + R\dot Q + \frac{Q}{C} = 0\;,
\end{equation}
which shows that an Ohmic resistor leads indeed to memoryless damping. These
considerations demonstrate that even without knowledge of the microscopic 
origin of dissipation in a resistor, we may employ the Ohmic spectral density 
(\ref{gli:eq:johmic}) to account for its dissipative nature.

The spectral density (\ref{gli:eq:johmic}) for Ohmic damping unfortunately 
diverges at high frequencies which, as already mentioned, cannot be the case
in practice. Even in theoretical considerations this feature of 
strictly Ohmic damping may lead to divergencies and a cutoff is needed for 
regularization. One possibility is the Drude cutoff,\index{Drude cutoff} where 
the spectral density\index{spectral density of bath oscillators!for Drude 
cutoff}\index{Drude cutoff!spectral density of bath oscillators for} 
\begin{equation}
J(\omega) = m\gamma\omega\frac{\omega_{\rm D}^2}{\omega^2+\omega_{\rm D}^2}
\label{gli:eq:jdrude}
\end{equation}
above frequencies of the order of $\omega_{\rm D}$ is suppressed. The 
corresponding damping kernel\index{damping kernel!for Drude cutoff}\index{Drude
cutoff!damping kernel for} reads
\begin{equation}
\gamma(t) = \gamma\omega_{\rm D}\exp(-\omega_{\rm D}\vert t\vert)\;.
\label{gli:eq:gdrude}
\end{equation}
This leads to memory effects in (\ref{gli:eq:effem}) for short times $t<
\omega_{\rm D}^{-1}$. For the long-time behaviour, however, only the Ohmic
low frequency behaviour of the spectral density (\ref{gli:eq:jdrude}) is 
relevant. If a Drude cutoff is introduced for technical reasons, the cutoff
frequency $\omega_{\rm D}$ should be much larger than all other frequencies 
appearing in the problem in order to avoid spurious effects.

The relation (\ref{gli:eq:j}) between the spectral density and the 
``microscopic'' parameters implies that one may set $c_n=m_n\omega_n^2$ without
loss of generality since the frequencies $\omega_n$ and the oscillator 
strengths $c_n^2/2m_n\omega_n$ can still be freely chosen. This special choice 
for the coupling constants has the advantage of a translationally invariant 
coupling \cite{gli:hakim85}
\begin{equation}
H=H_S + \sum_{n=1}^N\left(\frac{p_n^2}{2m_n} + \frac{m_n}{2}\omega_n^2
(x_n-q)^2\right)\;.
\end{equation}
Furthermore, we now can determine the total mass of environmental oscillators
\begin{equation}
\sum_{n=1}^N m_n =\frac{2}{\pi}\int_0^{\infty}\D \omega \frac{J(\omega)}
{\omega^3}\;.
\end{equation}
If the spectral density of bath oscillators at small frequencies takes the
form $J(\omega)\sim\omega^{\alpha}$, the total mass of bath oscillators is
infinite for $\alpha\le 2$. In particular, this includes the case of Ohmic
damping where a free damped particle executes a diffusive motion. In contrast,
for $\alpha>2$, the total mass is finite. In this case, the particle 
will behave for long times like it were free albeit possessing a renormalized 
mass due to the environmental coupling \cite{gli:grabe87}. We emphasize that 
the divergence of the total mass for $\alpha\le2$ is due to an infrared 
divergence and therefore independent of a high-frequency cutoff.

It is also useful to express the potential renormalization\index{potential
renormalization} introduced in (\ref{gli:eq:hsb}) in terms of the spectral 
density of bath oscillators. From (\ref{gli:eq:j}) it is straightforward to 
obtain
\begin{equation}
q^2\sum_{n=1}^N\frac{c_n^2}{2m_n\omega_n^2} =
\frac{q^2}{\pi}\int_0^{\infty}\D \omega\frac{J(\omega)}{\omega}\;.
\label{gli:eq:renorm}
\end{equation}
This term is infinite for strictly Ohmic damping but becomes finite when a 
high-frequency cutoff is introduced.

Finally, one finds for the noise correlation function\index{noise correlation
function} (\ref{gli:eq:zetazeta})
\begin{equation}
K(t)=\langle\zeta(t)\zeta(0)\rangle_{\rm B} =
\hbar\int_0^{\infty}\frac{\D \omega}{\pi}J(\omega)\left[
\coth\left(\frac{\hbar\beta\omega}{2}\right)\cos(\omega t) - \I 
\sin(\omega t)\right]\;.
\end{equation}
In the classical limit, $\hbar\to 0$, this correlation function reduces to
the real-valued expression
\begin{equation}
K(t) = m k_{\rm B}T\gamma(t)\;,
\label{gli:eq:clnoise}
\end{equation}
where we have made use of (\ref{gli:eq:gammaj}). For Ohmic damping this implies 
delta correlated, i.e.\ white, noise.  

In the quantum case, the noise correlation function is complex and can be
decomposed into its real and imaginary part
\begin{equation}
K(t)=K'(t)+\I K''(t)\;.
\end{equation}
Employing once more (\ref{gli:eq:gammaj}), one immediately finds that the 
imaginary part is related to the time derivative of the damping kernel by
\begin{equation}
K''(t) = \frac{m\hbar}{2}\frac{{\rm d}\gamma}{{\rm d}t}\;.
\end{equation}
For Ohmic damping, the real part reads
\begin{equation}
K'(t) = -\frac{\pi m\gamma}{(\hbar\beta)^2}\frac{1}{\displaystyle
\sinh^2\left(\frac{\pi t}{\hbar\beta}\right)}
\label{gli:eq:kreal}
\end{equation}
which implies that at zero temperature the noise is correlated even for long
times. The noise correlation then only decays algebraically like $1/t^2$ much 
in contrast to the classical result (\ref{gli:eq:clnoise}).

\subsection{Effective Action}
\label{gli:subsec:effact}

In the previous section we had eliminated the environmental degrees of
freedom to obtain the effective equation of motion (\ref{gli:eq:effem}) for the 
system degree of freedom alone. This section will be devoted to a discussion
of the corresponding procedure within the path integral 
formalism\index{elimination of environment|(}. 

We start to illustrate the basic idea by considering the time evolution of 
the full density matrix of system and environment
\begin{align}
W(q_{\rm f},x_{n\rm f},q'_{\rm f},x'_{n\rm f},t) &=
\int\D q_{\rm i}\D q'_{\rm i}\D x_{n\rm i}\D x'_{n\rm i}
K(q_{\rm f},x_{n\rm f},t,q_{\rm i},x_{n\rm i},0)\\
&\hspace{2truecm}\times W(q_{\rm i},x_{n\rm i},q'_{\rm i},x'_{n\rm  i},0)
K^*(q'_{\rm f},x'_{n\rm f},t,q'_{\rm i},x'_{n\rm i},0)\nonumber
\end{align}
which is induced by the two propagators $K$. Here, the coordinates $q$ and 
$x_n$ refer again to the system and bath degrees of freedom, respectively. The 
environment is assumed to be in thermal equilibrium described by the density 
matrix $W_{\beta}^{\rm B}$ while the system may be in a nonequilibrium state 
$\rho$. If we neglect initial correlations between system and environment, 
i.e.\ if we switch on the coupling after preparation of the initial state, the 
initial density matrix may be written in factorized form
\begin{equation}
W(q_{\rm i},x_{n\rm i},q'_{\rm i},x'_{n\rm i},0) = \rho(q_{\rm i},q'_{\rm i})
W_{\beta}^{\rm B}(x_{n\rm i},x'_{n\rm i})\;.
\end{equation}

Since we are only interested in the dynamics of the system degree of freedom, 
we trace out the environment. Then the time evolution may be expressed as
\begin{equation}
\rho(q_{\rm f},q'_{\rm f},t) = \int\D q_{\rm i}\D q'_{\rm i}
J(q_{\rm f},q'_{\rm f},t,q_{\rm i},q'_{\rm i},0) \rho(q_{\rm i},q'_{\rm i})
\end{equation}
with the propagating function
\begin{align}
\label{gli:eq:jprop}
J(q_{\rm f},q'_{\rm f},t,q_{\rm i},q'_{\rm i},0) =& \int\D x_{n\rm f}
\D x_{n\rm i}\D x'_{n\rm i}
K(q_{\rm f},x_{n\rm f},t,q_{\rm i},x_{n\rm i},0)\\
&\hspace{2truecm}\times W^{\rm B}_{\beta}(x_{n\rm i},x'_{n\rm i})
K^*(q'_{\rm f},x_{n\rm f},t,q'_{\rm i},x'_{n\rm i},0)\;.\nonumber
\end{align}
Here, the trace has been performed by setting $x_{n\rm f}=x'_{n\rm f}$ and
integrating over these coordinates. The propagators may be expressed as real 
time path integrals while the equilibrium density matrix of the bath is given 
by a path integral in imaginary time. Performing the path integrals and the 
conventional integrals appearing in (\ref{gli:eq:jprop}) one finds a functional 
depending on the system path. The important point is that this functional 
contains all information about the environment required to determine the system 
dynamics.

For factorizing initial conditions, the propagating function $J$ has been
calculated by Feynman and Vernon \cite{gli:feynm63} on the basis of the
Hamiltonian (\ref{gli:eq:clham}). More general initial conditions taking into 
account correlations between system and environment may be considered as well 
\cite{gli:grabe88}.

Instead of deriving the propagating function we will demonstrate how to 
trace the environment out of the equilibrium density matrix of system plus 
environment. While this task is conceptually similar and leads to the same 
physical insight, it is considerably less tedious. 

We start from the imaginary time path integral representation of the full
equilibrium density matrix
\begin{equation}
W_{\beta}(q,x_n,q',x'_n) = \frac{1}{\mathcal{Z}_{\beta}}\int{\cal D}\bar q
\left(\prod_{n=1}^N{\cal D}\bar x_n\right)\exp\!\left(-\frac{1}{\hbar}S^{\rm E}
[\bar q,\bar x_n]\right)
\end{equation}
where the paths run from $\bar q(0)=q'$ and $\bar x_n(0)=x'_n$ to $\bar 
q(\hbar\beta)=q$ and $\bar x_n(\hbar\beta)=x_n$. The Euclidean action
corresponding to the model Hamiltonian (\ref{gli:eq:clham}) reads in imaginary 
time
\begin{equation}
S^{\rm E}[\bar q,\bar x_n] = S_{\rm S}^{\rm E}[\bar q]
+ S_{\rm B}^{\rm E}[\bar x_n] + S_{\rm SB}^{\rm E}[\bar q, \bar x_n]
\end{equation}
with
\begin{align}
S^{\rm E}_{\rm S}[\bar q] &= \int_0^{\hbar\beta}\D \tau \left(
\frac{m}{2}\dot{\bar q}^2 + V(\bar q)\right)\\
S^{\rm E}_{\rm B}[\bar x_n] &= \int_0^{\hbar\beta}\D \tau \sum_{n=1}^N
\frac{m_n}{2}\left(\dot{\bar x}_n^2 + \omega_n^2\bar x_n^2\right)\\
S^{\rm E}_{\rm SB}[\bar q,\bar x_n] &= \int_0^{\hbar\beta}\D \tau
\left(-\bar q\sum_{n=1}^N c_n\bar x_n + \bar q^2\sum_{n=1}^N
\frac{c_n^2}{2m_n\omega_n^2}\right)\;.
\end{align}

The reduced density matrix of the system is obtained by tracing over the
environmental degrees of freedom
\begin{equation}
\begin{aligned}
\rho_{\beta}(q,q')&= \textrm{Tr}_{\rm B}\big(W_{\beta}(q,x_n,q',x'_n)\big)\\
&= \frac{1}{\mathcal{Z}_{\beta}}\int{\cal D}\bar q\int\prod_{n=1}^N\D x_n
\oint\prod_{n=1}^N{\cal D}\bar x_n \exp\!\left(-\frac{1}{\hbar}S^{\rm E}
[\bar q, \bar x_n]\right)
\end{aligned}
\end{equation}
where the circle on the second functional integral sign indicates that one has
to integrate over closed paths $\bar x_n(0)=\bar x_n(\hbar\beta)=x_n$ when
performing the trace. The dependence on the environmental coupling may be made
explicit by writing
\begin{equation}
\rho_{\beta}(q,q') = \frac{1}{\mathcal Z}\int{\cal D}\bar q 
\exp\!\left(-\frac{1}{\hbar}S_{\rm S}^{\rm E}[\bar q]\right){\cal F}[\bar q]
\label{gli:eq:rhosf}
\end{equation}
where the influence functional\index{influence functional} ${\cal F}[\bar q]$ 
describes the influence of the environment on the system. Here, the partition 
function $\mathcal Z$ should not be confused with the partition function 
$\mathcal Z_{\beta}$ of system plus environment. The relation between the two 
quantities will be discussed shortly.

Since the bath oscillators are not coupled among
each other, the influence functional may be decomposed into factors
corresponding to the individual bath oscillators
\begin{equation}
{\cal F}[\bar q] = \prod_{n=1}^N\frac{1}{\mathcal{Z}_n}{\cal F}_n[\bar q]
\label{gli:eq:inflfunc}
\end{equation}
where 
\begin{equation}
\mathcal{Z}_n=\frac{1}{2\sinh\left(\hbar\beta\omega_n/2\right)}
\label{gli:eq:zn}
\end{equation}
is the partition function of a single bath oscillator. The influence functional
of a bath oscillator can be expressed as
\begin{equation}
{\cal F}_n[\bar q] =\int\D x_n\oint{\cal D}\bar x_n 
\exp\!\left(-\frac{1}{\hbar}S^{\rm E}_n[\bar q,\bar x_n]\right)
\label{gli:eq:infln}
\end{equation}
with the action
\begin{equation}
S^{\rm E}_n[\bar q,\bar x_n] = \int_0^{\hbar\beta}\D \tau \frac{m_n}{2}
\left[\dot{\bar x}_n^2 + \omega_n^2\left(\bar x_n -\frac{c_n}{m_n\omega_n^2}
\bar q\right)^2\right]\;.
\label{gli:eq:actn}
\end{equation}
The partition function $\mathcal Z$ of the damped system is related to the full
partition function $\mathcal{Z}_{\beta}$ by the partition function of the 
environmental oscillators $\mathcal{Z}_{\rm B}=\prod_{n=1}^N \mathcal{Z}_n$ 
according to $\mathcal{Z}=\mathcal{Z}_{\beta}/\mathcal{Z}_{\rm B}$. In the 
limit of vanishing coupling, $c_n=0$, the influence functional becomes 
${\cal F}[\bar q] =1$ so that (\ref{gli:eq:rhosf}) reduces to the path integral
representation of the density matrix of an isolated system as it should.

Apart from the potential renormalization term proportional to $\bar q^2$, the 
action (\ref{gli:eq:actn}) describes a driven harmonic oscillator. We may 
therefore make use of our results from Sect.~\ref{gli:subsec:dho}. After 
analytic continuation $t\to -\I \hbar\beta$ in (\ref{gli:eq:dhoaction})
and setting $x_{\rm i}=x_{\rm f}=x_n$ one finds for the classical Euclidean 
action 
\begin{align}
S^{\rm E,cl}_n[\bar q] &= m_n\omega_n\frac{\cosh(\hbar\beta\omega_n)-1}
{\sinh(\hbar\beta\omega_n)}x_n^2 \nonumber\\
&\qquad- c_n\int_0^{\hbar\beta}\!\!\D \tau
\frac{\sinh(\omega_n\tau)+\sinh(\omega_n(\hbar\beta-\tau))}
{\sinh(\hbar\beta\omega_n)}x_n\bar q(\tau)\nonumber\\
&\qquad-\frac{c_n^2}{m_n\omega_n}\int_0^{\hbar\beta}\!\!\D \tau
\int_0^{\tau}\!\!\D \sigma \frac{\sinh(\omega_n(\hbar\beta-\tau))
\sinh(\omega_n\sigma)}{\sinh(\hbar\beta\omega_n)}\bar q(\tau)\bar q(\sigma)
\nonumber\\
&\qquad + \frac{c_n^2}{2m_n\omega_n^2}\int_0^{\hbar\beta}\!\!\D \tau \bar
q^2(\tau)\;.
\label{gli:eq:seclocs}
\end{align}
In view of the required integration over $x_n$ one completes the square 
\begin{align}
S^{\rm E,cl}_n[\bar q] &= m_n\omega_n\frac{\cosh(\hbar\beta\omega_n)-1}
{\sinh(\hbar\beta\omega_n)}(x_n-x_n^{(0)})^2 - 
\int_0^{\hbar\beta}\!\!\D \tau\int_0^{\tau}\!\!\D \sigma 
K_n(\tau-\sigma)\bar q(\tau)\bar q(\sigma)\nonumber\\ 
&\qquad+ \frac{c_n^2}{2m_n\omega_n^2}\int_0^{\hbar\beta}\!\!\D \tau \bar
q^2(\tau)
\label{gli:eq:secl}
\end{align}
where $x_n^{(0)}$ does not need to be specified since it drops out after
integration.

The integral kernel appearing in (\ref{gli:eq:secl}) follows from
(\ref{gli:eq:seclocs}) as
\begin{equation}
K_n(\tau)=\frac{c_n^2}{2m_n\omega_n}\frac{\cosh\left(\omega_n\big(
\dfrac{\hbar\beta}{2}-\tau\big)\right)}{\sinh\left(\dfrac{\hbar\beta\omega_n}
{2}\right)}=K_n(\hbar\beta-\tau)
\label{gli:eq:knkernel}
\end{equation}
and therefore can be identified as the noise correlation function 
(\ref{gli:eq:noisecorr}) in imaginary time\index{noise correlation function}
\begin{equation}
K_n(\tau)=\frac{1}{\hbar}\langle\zeta_n(-\I \tau)
\zeta_n(0)\rangle_{\rm B}\;.
\label{gli:eq:kntau}
\end{equation}
The term in (\ref{gli:eq:secl}) containing this kernel is quite unusual for
an action. The double integral describes a nonlocal contribution where the 
system trajectory interacts with itself. This self-interaction is mediated by 
the environment as can be seen from the factor $c_n^2$ in 
(\ref{gli:eq:knkernel}).

The integral kernel $K_n(\tau)$ is only needed in an interval of length
$\hbar\beta$. Periodic continuation outside of this interval therefore allows 
us to expand the kernel into a Fourier series
\begin{align}
K_n(\tau) &= \frac{c_n^2}{\hbar\beta m_n\omega_n}\sum_{l=-\infty}^{\infty}
\frac{\omega_n}{\omega_n^2+\nu_l^2}\exp(\I \nu_l\tau)\nonumber\\
&= \frac{c_n^2}{\hbar\beta m_n\omega_n^2}\sum_{l=-\infty}^{\infty}
\exp(\I \nu_l\tau) -\frac{c_n^2}{\hbar\beta m_n\omega_n^2}
\sum_{l=-\infty}^{\infty}\frac{\nu_l^2}{\omega_n^2+\nu_l^2}\exp(\I 
\nu_l\tau)\nonumber\\
&= \frac{c_n^2}{m_n\omega_n^2}\sum_{j=-\infty}^{\infty}\delta(\tau-
j\hbar\beta) -k_n(\tau)\;,
\label{gli:eq:ksplit}
\end{align}
where the Matsubara frequencies $\nu_l$ have been defined in 
(\ref{gli:eq:matsubara}). In (\ref{gli:eq:ksplit}), we split the kernel into 
two parts. The first term contains delta functions which lead to a local 
contribution to the action. Noting that due to the region of integration in 
(\ref{gli:eq:secl}) only half of the delta function contributes, this local 
term just cancels the potential renormalization\index{potential 
renormalization} (\ref{gli:eq:renorm}). We are therefore left with the nonlocal 
kernel
\begin{equation}
k_n(\tau)=\frac{c_n^2}{\hbar\beta m_n\omega_n^2}\sum_{l=-\infty}^{\infty}
\frac{\nu_l^2}{\omega_n^2+\nu_l^2}\exp(\I \nu_l\tau)\;.
\end{equation}
It can be shown that this kernel no longer contains a local contribution by
writing
\begin{align}
&\int_0^{\hbar\beta}\!\!\D \tau\int_0^{\tau}\!\!\D \sigma 
k_n(\tau-\sigma) \bar q(\tau)\bar q(\sigma)\\ 
&\qquad\qquad= -\frac{1}{2}\int_0^{\hbar\beta}\!\!\D \tau\int_0^{\tau}
\!\!\D \sigma k_n(\tau-\sigma)\left[\big(\bar q(\tau)-\bar q(\sigma)
\big)^2-\big(\bar q(\tau)^2+\bar q(\sigma)^2\big)\right]\;.\nonumber
\end{align}
The first term is manifestly nonlocal because it contains the difference $\bar
q(\tau)-\bar q(\sigma)$. Exploiting the symmetry of $k_n(\tau)$, the second 
term can be expressed as
\begin{equation}
\frac{1}{2}\int_0^{\hbar\beta}\!\!\D \tau\int_0^{\tau}\!\!
\D \sigma k_n(\tau-\sigma)\left(\bar q(\tau)^2+\bar q(\sigma)^2\right)
= \int_0^{\hbar\beta}\!\!\D \tau \bar q(\tau)^2\int_0^{\hbar\beta}\!\!
\D \sigma k_n(\sigma)\;.
\end{equation}
Therefore, this term potentially could result in a local contribution. However, 
the time integral over the interval from 0 to $\hbar\beta$ corresponds to the 
$l=0$ Fourier component which vanishes for $k_n(\tau)$. As a consequence, the 
kernel $k_n(\tau)$ indeed gives rise to a purely nonlocal contribution to the 
action.

Now, we can carry out the Gaussian integral over $x_n$ appearing in the 
influence functional (\ref{gli:eq:infln}). With the action (\ref{gli:eq:secl}) 
we find 
\begin{equation}
{\cal F}_n[\bar q]=\mathcal{Z}_n \exp\!\left(-\frac{1}{2\hbar}
\int_0^{\hbar\beta}
\D \tau\int_0^{\hbar\beta}\D \sigma k_n(\tau-\sigma)\bar q(\tau)
\bar q(\sigma)\right)\;.
\end{equation}
The partition function $\mathcal{Z}_n$ arises from the fluctuation contribution 
and may be shown to be given by (\ref{gli:eq:zn}) for example by comparison 
with the uncoupled case $c_n=0$\index{elimination of environment|)}.

With (\ref{gli:eq:inflfunc}) we finally obtain the influence 
functional\index{influence functional}
\begin{equation}
{\cal F}[\bar q] = \exp\!\left(-\frac{1}{2\hbar}\int_0^{\hbar\beta}\D \tau
\int_0^{\hbar\beta}\D \sigma k(\tau-\sigma)\bar q(\tau)\bar q(\sigma)\right)
\end{equation}
with
\begin{align}
k(\tau)=\sum_{n=1}^N k_n(\tau) &=
\sum_{n=1}^N\frac{c_n^2}{\hbar\beta m_n\omega_n^2}\sum_{l=-\infty}^{\infty}
\frac{\nu_l^2}{\omega_n^2+\nu_l^2}\exp(\I \nu_l\tau)\nonumber\\
&=\frac{2}{\hbar\beta}\int_0^{\infty}\frac{\D \omega}{\pi}
\frac{J(\omega)}{\omega}\sum_{l=-\infty}^{\infty}\frac{\nu_l^2}{\omega^2
+\nu_l^2}\exp(\I \nu_l\tau)
\label{gli:eq:ktaua}
\end{align}
where we have made use of the spectral density of bath oscillators 
(\ref{gli:eq:j}) to obtain the last line.

The kernel $k(\tau)$ may be related to the damping kernel $\gamma(t)$ by 
observing that the Laplace transform of the latter is given by
\begin{align}
\hat\gamma(z) = \int_0^{\infty}\D t\exp(-zt)\gamma(t)
&=\frac{2}{m}\int_0^{\infty}\frac{\D \omega}{\pi}\frac{J(\omega)}{\omega}
\int_0^{\infty}\D t\exp(-zt)\cos(\omega t)\nonumber\\
&=\frac{2}{m}\int_0^{\infty}\frac{\D \omega}{\pi}\frac{J(\omega)}{\omega}
\frac{z}{z^2+\omega^2}\;.
\label{gli:eq:gammalapl}
\end{align}
In the first line we have employed the relation (\ref{gli:eq:gammaj}) between 
the damping kernel and the spectral density of bath oscillators. In view of
(\ref{gli:eq:ktaua}) and (\ref{gli:eq:gammalapl}) we can finally express
the kernel as
\begin{equation}
k(\tau)=\frac{m}{\hbar\beta}\sum_{l=-\infty}^{\infty} \vert\nu_l\vert
\hat\gamma(\vert\nu_l\vert)\exp(\I \nu_l\tau)\;.
\label{gli:eq:ktau}
\end{equation}
For strictly Ohmic damping, this kernel is highly singular and we therefore 
introduce a Drude cutoff. The Laplace transform of the corresponding damping 
kernel is obtained from (\ref{gli:eq:gdrude}) as
\begin{equation}
\hat\gamma(z)=\frac{\gamma\omega_{\rm D}}{\omega_{\rm D}+z}
\label{gli:eq:drudelap}
\end{equation}
which reduces to $\hat\gamma(z)=\gamma$ for strictly Ohmic damping. Keeping
the leading terms in the cutoff frequency $\omega_{\rm D}$, the kernel now 
reads
\begin{equation}
k(\tau) = m\gamma\omega_{\rm D}\sum_{n=-\infty}^{\infty}\delta(\tau
-n\hbar\beta) - \frac{\pi m\gamma}{(\hbar\beta)^2}\frac{1}{\sin^2\left(
\dfrac{\pi\tau}{\hbar\beta}\right)} + O(\omega_{\rm D}^{-1})\;.
\label{gli:eq:ktauf}
\end{equation}
For low temperatures, this gives rise to a long range interaction between
different parts of the system trajectory. This reminds us of the algebraic
decay of the real part (\ref{gli:eq:kreal}) of the noise correlation 
function\index{noise correlation function} $K(t)$ and in fact it follows from 
our previous discussion that up to the periodic delta function appearing in 
(\ref{gli:eq:ktauf}) the kernel $k(\tau)$ equals $-K(-\I\tau)$.

Summarizing this calculation we obtain the important result that the influence 
of the environment on the system may be taken into account by adding a nonlocal 
contribution to the action. We then obtain the effective action\index{effective
action}\index{action!effective}
\begin{equation}
S_{\rm eff}^{\rm E}[\bar q] = S_{\rm S}^{\rm E}[\bar q] + \frac{1}{2}
\int_0^{\hbar\beta}\!\D \tau\int_0^{\hbar\beta}\!\D \sigma\,
k(\tau-\sigma) \bar q(\tau)\bar q(\sigma)
\label{gli:eq:effaction}
\end{equation}
with $k(\tau)$ given by (\ref{gli:eq:ktau}). The elimination of the environment
within the real time path integral formalism, e.g. along the lines of the
calculation by Feynman and Vernon \cite{gli:feynm63} mentioned at the beginning
of this section, would have led to an effective action of a structure similar
to (\ref{gli:eq:effaction}). An important difference consists in the fact that
the propagation of a density matrix involves two paths instead of one. In 
addition, the integral kernel then of course appears in its real time version.

\section{Damped Harmonic Oscillator}
\label{gli:sec:dho}\index{damped harmonic oscillator|(}

\subsection{Partition Function}
In this final section we will apply the results of the previous sections to
the damped harmonic oscillator where exact results may be obtained 
analytically. The Hamiltonian describing system and environment is given by 
(\ref{gli:eq:clham})--(\ref{gli:eq:hsb}) with the potential
\begin{equation}
V(q) = \frac{m}{2}\omega_0^2q^2\;.
\end{equation}
As we have seen in Sect.~\ref{gli:sec:model}, there is no point in dealing
with all the microscopic parameters present in the Hamiltonians 
(\ref{gli:eq:hb}) and (\ref{gli:eq:hsb}). Instead, it is sufficient to specify
the spectral density of bath oscillators (\ref{gli:eq:j}). In the following,
we will mostly assume Ohmic damping, i.e. $J(\omega)=m\gamma\omega$,
and introduce a high-frequency cutoff of the Drude type (\ref{gli:eq:jdrude})
when necessary.

In the general discussion of dissipative quantum systems we have concentrated
on imaginary time calculations and we will therefore try to take this approach
as a starting point of the following considerations. Probably the most 
important quantity which can be obtained in imaginary time is the partition
function which in statistical mechanics can be viewed as a generating function
for expectation values. Based on our previous discussion of the partition
function of the undamped harmonic oscillator in Sect.~\ref{gli:subsec:itpi}
and the effective action in imaginary time in Sect.~\ref{gli:subsec:effact},
it is rather straightforward to obtain the partition function of the damped
harmonic oscillator.

According to (\ref{gli:eq:partho}) and (\ref{gli:eq:sinhrel}) we can express
the partition function of the undamped harmonic oscillator as
\begin{equation}
\mathcal{Z}_{\rm u} = \frac{1}{\hbar\beta\omega_0}\prod_{n=1}^{\infty}
\frac{\nu_n^2}{\nu_n^2+\omega_0^2}\;.
\label{gli:eq:zuho}
\end{equation}
We remind the reader that the denominator of the product stems from the 
fluctuation determinant associated with the Euclidean action 
\begin{equation}
S^{\rm E}[q] = \int_0^{\hbar\beta}\D \tau\left(\frac{m}{2}\dot q^2 +
\frac{m}{2}\omega_0^2 q^2\right)\;.
\end{equation}
As we have seen in Sect.~\ref{gli:subsec:effact}, the coupling to the 
environment leads to an additional, nonlocal term to the action. Expanding the 
fluctuations in a Fourier series as we did on p.~\pageref{gli:page:zfou} and 
making use of the Fourier decomposition (\ref{gli:eq:ktau}) of the integral 
kernel $k(\tau)$, we conclude that an additional term $\nu_n\hat\gamma(\nu_n)$ 
appears in the fluctuation determinant. Modifying (\ref{gli:eq:zuho}) 
accordingly, we find for the partition function of the damped harmonic
oscillator\index{partition function!of damped harmonic oscillator}\index{damped
harmonic oscillator!partition function of}
\begin{equation}
\mathcal{Z} = \frac{1}{\hbar\beta\omega_0}\prod_{n=1}^{\infty}
\frac{\nu_n^2}{\nu_n^2+\nu_n\hat\gamma(\nu_n)+\omega_0^2}\;.
\label{gli:eq:partfunc}
\end{equation}
For strictly Ohmic damping, we have $\hat\gamma(\nu_n)=\gamma$. Since infinite
products over terms of the form $1+a/n$ for large $n$ do not converge, we are 
forced to introduce a high-frequency cutoff in order to obtain a finite result. 
One possibility is the Drude cutoff (\ref{gli:eq:gdrude}) with $\hat\gamma$ 
given by (\ref{gli:eq:drudelap}).

In the following section we will try to extract some interesting information 
from the partition function and in the process will get an idea of where the
difficulties for strictly Ohmic damping arise from.

\subsection{Ground State Energy and Density of States}

A thermodynamic quantity directly related to the partition function is the free 
energy which can be obtained from the former by means of
\begin{equation}
F = -\frac{1}{\beta}\ln(\mathcal{Z})\;.
\end{equation}
In the limit of zero temperature, the free energy becomes the ground state
energy of the undamped oscillator shifted due to the coupling to the 
environment. For the free energy\index{free energy of damped harmonic 
oscillator}\index{damped harmonic oscillator!free energy of}, we find with 
(\ref{gli:eq:partfunc})
\begin{equation}
F = \frac{1}{\beta}\ln(\hbar\beta\omega_0) + \frac{1}{\beta}\sum_{n=1}^{\infty}
\ln\!\left(1+\frac{\hat\gamma(\nu_n)}{\nu_n}+
\frac{\omega_0^2}{\nu_n^2}\right)\;.
\end{equation}
In the limit $\beta\to\infty$ the spacing between the Matsubara frequencies
$\nu_n$ goes to zero and the sum turns into the ground state energy of the
damped oscillator given by the integral\index{damped harmonic oscillator!ground
state energy of}
\begin{equation}
\varepsilon_0 = \frac{\hbar}{2\pi}\int_0^{\infty}\D \nu\ln\!\left(
1+\frac{\hat\gamma(\nu)}{\nu}+\frac{\omega_0^2}{\nu^2}\right)\;.
\label{gli:eq:e0}
\end{equation}

It is particularly interesting to consider the case of weak coupling where 
connection can be made to results of perturbation theory. This will also help
to understand the physical meaning of a ground state energy of a dissipative
system derived from a free energy. An expansion of (\ref{gli:eq:e0}) including 
terms of order $\gamma$ yields
\begin{equation}
\varepsilon_0 = \frac{\hbar}{2\pi}\int_0^{\infty}\D \nu \ln\!\left(1+
\frac{\omega_0^2}{\nu^2}\right) +\frac{\hbar}{2\pi}\int_0^{\infty}\D \nu
\frac{\nu}{\nu^2+\omega_0^2}\hat\gamma(\nu)\;.
\label{gli:eq:e0a}
\end{equation}
Evaluation of the first integral yields the expected result $\hbar\omega_0/2$, 
i.e. the ground state energy of the undamped harmonic oscillator. The second 
integral represents the shift due to the coupling to the environmental 
oscillators and can be expressed in terms of the spectral density of bath 
oscillators $J(\omega)$. Recalling (\ref{gli:eq:gammalapl}) one can perform the 
integral over $\nu$ and the ground state energy (\ref{gli:eq:e0a}) in the 
presence of damping becomes
\begin{equation}
\varepsilon_0 = \frac{\hbar\omega_0}{2} + \frac{\hbar}{2\pi m}\int_0^{\infty}
\D \omega J(\omega)\frac{1}{\omega(\omega_0+\omega)}\;.
\end{equation}
In order to facilitate the physical interpretation, we rewrite this result as
\begin{equation}
\varepsilon_0 = \frac{\hbar\omega_0}{2} - \frac{\hbar}{2\pi m\omega_0}
\int_0^{\infty}\D \omega J(\omega)\left(\frac{1}{\omega_0+\omega}-
\frac{1}{\omega}\right)\;.
\label{gli:eq:e0b}
\end{equation}

\begin{figure}[t]
\begin{center}
\includegraphics[width=0.3\textwidth]{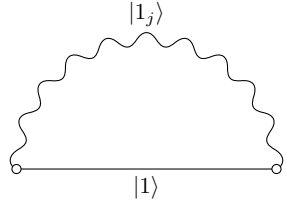}
\end{center}
\caption{The ground state energy of the harmonic oscillator is shifted by
a transition to the first excited state accompanied by a virtual excitation
of the $j$-th environmental mode}
\label{gli:fig:lambshift}
\end{figure}

The first term of order $\gamma$ may be interpreted in analogy to the Lamb 
shift\index{Lamb shift}. There, an atomic level is shifted by creation and 
subsequent annihilation of a virtual photon as a consequence of the coupling to 
the electromagnetic vacuum. In our case, the atomic level is replaced by the
ground state of the harmonic oscillator and the environmental oscillators are
completely equivalent to the modes of the electromagnetic field. The pictorial
representation of this process is shown in Fig.~\ref{gli:fig:lambshift}. The
coupling Hamiltonian (\ref{gli:eq:hsb}) allows a transition from the 
ground state into the first excited state $\vert 1\rangle$ by excitation of the
$j$-th environmental oscillator into its first excited state $\vert 
1_j\rangle$.

The energy shift associated with the diagram depicted in 
Fig.~\ref{gli:fig:lambshift} is given by second order perturbation theory as
\begin{equation}
\Delta_0 = \sum_{j=1}^N\frac{\vert\langle 1,1_j\vert c_jqx_j
\vert 0,0\rangle\vert^2}{-\hbar\omega_0-\hbar\omega_j}
\label{gli:eq:d0}
\end{equation}
where the denominator is determined by the energy of the intermediate state
$\vert 1,1_j\rangle$. With the matrix element
\begin{equation}
\langle 1,1_j\vert qx_j\vert 0,0\rangle = \frac{\hbar}
{2(mm_j\omega_0\omega_j)^{1/2}}
\end{equation}
and the relation (\ref{gli:eq:j}) for the spectral density of bath oscillators
we get
\begin{equation}
\Delta_0 = -\frac{\hbar}{2\pi m\omega_0}\int_0^{\infty}\D \omega J(\omega)
\frac{1}{\omega_0+\omega}
\label{gli:eq:d0a}
\end{equation}
which is just the first term of order $\gamma$ in (\ref{gli:eq:e0b}). As we 
know, the bilinear coupling Hamiltonian appearing in (\ref{gli:eq:d0}) gives
rise to a renormalization of the potential which has been taken care of by
the second term in the Hamiltonian (\ref{gli:eq:hsb}). The result 
(\ref{gli:eq:d0a}) contains this potential renormalization\index{potential
renormalization} since only the bilinear coupling term has been considered. In 
(\ref{gli:eq:e0b}), which results from the full Hamiltonian, this effect is 
subtracted off by the second term under the integral in (\ref{gli:eq:e0b}) as 
can be verified by comparison with (\ref{gli:eq:renorm}).

It is obvious that for strictly Ohmic damping with $J(\omega)=m\gamma\omega$
the correction (\ref{gli:eq:d0a}) and with it the ground state energy 
(\ref{gli:eq:e0b}) will display an ultraviolet divergence which is due to the 
unphysical behaviour of the spectral density $J(\omega)$ at large frequencies. 
Assuming a Drude cutoff we find with (\ref{gli:eq:jdrude}) to leading order in 
the cutoff frequency $\omega_{\rm D}$ the finite result
\begin{equation}
\Delta_0 = -\frac{\hbar\gamma\omega_{\rm D}}{4\omega_0} + \frac{\hbar\gamma}
{2\pi}\ln\!\left(\frac{\omega_{\rm D}}{\omega_0}\right)+
O(\omega_{\rm D}^{-1})\;.
\label{gli:eq:d0b}
\end{equation}
The negative first term corresponds to the potential 
renormalization\index{potential renormalization} which is no longer present in 
the ground state energy $\varepsilon_0$. The second term, on the other hand, is 
positive and thus leads to an increase of the ground state energy. 

Not only the ground state energy can be derived from the partition function
but one may also formally introduce a density of states $\rho(E)$ of the damped 
system according to \cite{gli:hanke95}\index{density of states!of damped 
harmonic oscillator}\index{damped harmonic oscillator!density of states of}
\begin{equation}
\mathcal{Z}(\beta) = \int_0^{\infty}\D E\rho(E)\exp(-\beta E)\;.
\label{gli:eq:zbeta}
\end{equation}
Inversion of the Laplace transformation allows to determine $\rho(E)$ from the 
partition function according to
\begin{equation}
\rho(E) = \frac{1}{2\pi\I }\int_{c-{\rm i}\infty}^{c+{\rm i}\infty}
\D \beta\mathcal{Z}(\beta)\exp(\beta E)
\label{gli:eq:rhoinvlap}
\end{equation}
where the constant $c$ has to be chosen such that the line of integration is
to the right of all poles of $\mathcal{Z}(\beta)$. 

Once a high-frequency cutoff for the spectral density of bath oscillators is
specified, the inverse Laplace transform in (\ref{gli:eq:rhoinvlap}) may be 
evaluated either numerically or by contour integration. The second approach 
leads to a series which again has to be evaluated numerically
\cite{gli:hanke95}. However, it is not necessary to introduce a cutoff provided 
we shift the energy by the ground state energy $\varepsilon_0$ which in fact is 
the only divergent quantity in this problem. Such a shift may be performed by 
considering $\mathcal{Z}\exp(\beta\varepsilon_0)$ instead of $\mathcal{Z}$ 
itself. To demonstrate that this procedure renders the cutoff irrelevant, we 
will restrict ourselves to the limit of weak damping and large cutoff 
considered before even though a more general treatment is feasible. 

In a first step we decompose the infinite product appearing in the partition
function (\ref{gli:eq:partfunc}) with Drude cutoff (\ref{gli:eq:drudelap})
into a factor where the limit $\omega_{\rm D}\to\infty$ can safely be taken
and a factor still containing the cutoff frequency 
\begin{equation}
\mathcal{Z} = \frac{1}{\hbar\beta\omega_0}\prod_{n=1}^{\infty}
\frac{\nu_n^2+\gamma\nu_n}{\nu_n^2+\gamma\nu_n+\omega_0^2}\prod_{n=1}^{\infty}
\frac{1}{\displaystyle 1+\frac{\gamma\omega_{\rm D}}
{\nu_n(\nu_n+\omega_{\rm D})}}\;.
\end{equation}
It is the last product which has to be analyzed with care because it vanishes
in the limit $\omega_D\to\infty$. To leading order in $\gamma$ one finds
\begin{equation}
\begin{split}
\ln\!\left(\prod_{n=1}^{\infty} 1+\frac{\gamma\omega_{\rm D}}{\nu_n(\nu_n+
\omega_{\rm D})}\right) &= \sum_{n=1}^{\infty}\frac{\gamma\omega_{\rm D}}
{\nu_n(\nu_n+\omega_{\rm D})}\\
&=\frac{\hbar\beta\gamma}{2\pi}\psi\!\left(1+\frac{\hbar\beta\omega_{\rm D}}
{2\pi}\right)
\end{split}
\end{equation}
where we have introduced the digamma function \cite{gli:abram72}
\begin{equation}
\psi(1+z) = -\mathcal{C}+\sum_{n=1}^{\infty}\frac{z}{n(n+z)}\;.
\end{equation}
Here, $\mathcal{C}=0.577\dots$ is the Euler constant. With the leading 
asymptotic behaviour $\psi(1+z)\sim\ln(z)$ for large arguments $z$, the 
partition function for large cutoff frequency becomes
\begin{equation}
\mathcal{Z} = \frac{1}{\hbar\beta\omega_0}\prod_{n=1}^{\infty}
\frac{\nu_n^2+\gamma\nu_n}{\nu_n^2+\gamma\nu_n+\omega_0^2}
\left(\frac{\hbar\beta\omega_{\rm D}}{2\pi}\right)^{-\hbar\beta\gamma/2\pi}\;.
\end{equation}
On the other hand, we know from our result (\ref{gli:eq:d0b}) that the ground
state energy diverges to leading order with $(\hbar\gamma/2\pi)\ln(\omega_{\rm
D})$. Therefore, multiplication of the partition function with $\exp(\beta
\varepsilon_0)$ leads indeed to an expression with a finite value in the limit
of infinite cutoff frequency.

\begin{figure}[t]
\begin{center}
\includegraphics[width=0.5\textwidth]{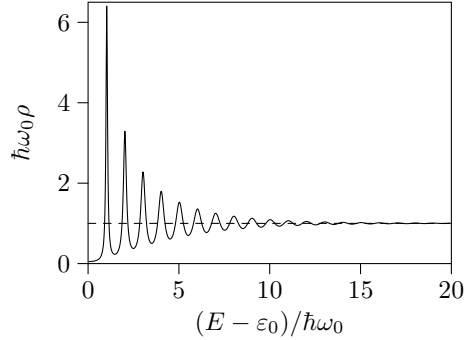}
\end{center}
\caption{The density of states (\protect\ref{gli:eq:rhoinvlap})\index{density 
of states!of damped harmonic oscillator}\index{damped harmonic 
oscillator!density of states of} of the damped harmonic oscillator is shown for 
$\gamma/2\omega_0=0.05$. A delta function contribution at $E=\varepsilon_0$ has 
been omitted. The dashed line marks the average density of states 
$1/\hbar\omega_0$}
\label{gli:fig:rho}
\end{figure}

Before taking a look at numerical results, we remark that the partition 
function contains a pole at $\beta=0$ with residue $1/\hbar\omega_0$. This
pole represents the Laplace transform of the constant $1/\hbar\omega_0$ and
therefore is related to the average density of states which takes the value
of the undamped case where the energy spacing between adjacent levels is
$\hbar\omega_0$. In Fig.~\ref{gli:fig:rho} we present the density of states for 
weak damping, $\gamma/2\omega_0=0.05$. A delta function contribution at 
$E=\varepsilon_0$ has been omitted. Due to the weak damping we find well 
defined peaks which are close to the energies expected for an undamped 
oscillator. With increasing energy the levels become broader. For stronger 
damping, only the lowest levels can be resolved and a level shift induced by 
the damping becomes visible. 

The behaviour of the level widths shown in Fig.~\ref{gli:fig:rho} is 
consistent with the result of a perturbative treatment. According to Fermi's
golden rule, the width of the $n$-th level is given by
\begin{equation}
\begin{split}
\Gamma_n &= \frac{2\pi}{\hbar^2}\sum_{j=1}^{\infty}\left[
\big\vert\langle n+1,1_j\vert c_jqx_j\vert n,0\rangle\big\vert^2
\delta(-\omega_0-\omega_j)\right.\\
&\qquad\qquad\quad+
\left.\big\vert\langle n-1,1_j\vert c_jqx_j\vert n,0\rangle\big\vert^2
\delta(\omega_0-\omega_j)\right]
\end{split}
\end{equation}
where we have already taken into account that the matrix element of the
dipole-type coupling only connects the state $n$ to its nearest neighbours.
Because of energy conservation the first part of the sum never contributes.
With the matrix elements
\begin{equation}
\langle n-1,1_j\vert qx_j\vert n,0\rangle = \frac{\hbar}
{2(mm_j\omega_0\omega_j)^{1/2}}n^{1/2}
\end{equation}
we thus find for the width of the $n$-th level in terms of the spectral 
density of bath oscillators\index{level width for damped harmonic 
oscillator}\index{damped harmonic oscillator!level width for}
\begin{equation}
\Gamma_n = \frac{n}{m\omega_0}J(\omega_0)\;.
\end{equation}
For Ohmic damping, $J(\omega)=m\gamma\omega$, we finally get
\begin{equation}
\Gamma_n = n\gamma\;.
\end{equation}
As observed before, the levels broaden with increasing damping strength 
$\gamma$ and level number $n$. We remark that one can demonstrate by a 
semiclassical analysis of other one-dimensional potentials that it is indeed 
the level number and not the energy which is decisive for the level width 
\cite{gli:ingol01}.

\subsection{Position Autocorrelation Function}

In the introduction we have mentioned fluctuations as one of the effects 
arising from the coupling to an environment. Even if a system is in thermal
equilibrium with its environment, fluctuations due to the noise term 
(\ref{gli:eq:noise}) will be present. The appropriate quantity to describe this 
phenomenon are equilibrium correlation functions like the position 
autocorrelation function\index{position autocorrelation function}
\begin{equation}
C(t) = \langle q(t)q(0)\rangle = \mathrm{Tr}(q(t)q(0)\rho_{\beta})\;.
\end{equation}
From this quantity one can derive all other equilibrium correlation functions
of the damped harmonic oscillator as is discussed in \cite{gli:grabe84}.

We now want to determine this correlation function by first calculating its
imaginary time version and start with the Euclidean action
\begin{equation}
\begin{split}
S^{\rm E}[q] &= \int_0^{\hbar\beta}\!\D \tau\left(\frac{m}{2}\dot q^2
+\frac{m}{2}\omega_0^2q^2\right)+\frac{1}{2}\int_0^{\hbar\beta}\!\!\D \tau
\int_0^{\hbar\beta}\!\!\D \sigma k(\tau-\sigma)q(\tau)q(\sigma)\\
&\qquad\qquad+\int_0^{\hbar\beta}\!\D \tau F(\tau)q(\tau)\;.
\label{gli:eq:seho}
\end{split}
\end{equation}
The second term accounts for the coupling to the environment as we have shown
in Sect.~\ref{gli:subsec:effact}. In addition, we have included the third term
corresponding to an external force in imaginary time. This constitutes a useful
trick which will allow us to determine the correlation function by variation
with respect to this force
\begin{equation}
\langle q(\tau)q(\sigma)\rangle = \hbar^2\mathrm{Tr}\!\left.\left(
\frac{\delta}{\delta F(\tau)}\frac{\delta}{\delta F(\sigma)}\rho\right)
\right\vert_{F=0}\;.
\label{gli:eq:varqq}
\end{equation}

As we know already from Sect.~\ref{gli:subsec:dho}, the force does not appear
in the fluctuation part. It is therefore sufficient, to restrict our
attention to the classical path. The classical equation of motion following
from variation of the action (\ref{gli:eq:seho})
\begin{equation}
m\ddot q(\tau) - \int_0^{\hbar\beta}\!\D \sigma k(\tau-\sigma)q(\sigma)
-m\omega_0^2q(\tau) = F(\tau)
\end{equation}
is most conveniently solved by Fourier transformation on the interval from 0
to $\hbar\beta$. Introducing the Fourier transforms
\begin{equation}
q(\tau) = \frac{1}{\hbar\beta}\sum_{n=-\infty}^{\infty}q_n\exp(\I \nu_n
\tau)
\end{equation}
and
\begin{equation}
F(\tau) = \frac{m}{\hbar\beta}\sum_{n=-\infty}^{\infty}f_n\exp(\I \nu_n
\tau)
\end{equation}
and making use of the Fourier representation (\ref{gli:eq:ktau}) of $k(\tau)$ 
for Ohmic damping, we find for the Fourier coefficients of the classical 
solution
\begin{equation}
q_n^{\rm cl} = - \frac{f_n}{\nu_n^2+\gamma\vert\nu_n\vert+\omega_0^2}\;.
\label{gli:eq:clsit}
\end{equation}
Inserting this result into the Fourier representation of the action
(\ref{gli:eq:seho}) 
\begin{equation}
S^{\rm E}[q] = \frac{m}{2\hbar\beta}\sum_{n=-\infty}^{\infty}
\big[(\nu_n^2+\gamma\vert\nu_n\vert+\omega_0^2)q_nq_{-n}
+f_nq_{-n}+f_{-n}q_n\big]
\end{equation}
yields the classical Euclidean action
\begin{equation}
S^{\rm E}_{\rm cl} = -\frac{m}{2\hbar\beta}\sum_{n=-\infty}^{\infty}
\frac{f_nf_{-n}}{\nu_n^2+\gamma\vert\nu_n\vert+\omega_0^2}
\end{equation}
or equivalently
\begin{equation}
\begin{aligned}
S^{\rm E}_{\rm cl} &= -\frac{1}{2m\hbar\beta}\sum_{n=-\infty}^{\infty}
\frac{1}{\nu_n^2+\gamma\vert\nu_n\vert+\omega_0^2}\\
&\hspace{3truecm}\times\int_0^{\hbar\beta}\!\!
\D \tau\int_0^{\hbar\beta}\!\!\D \sigma F(\tau)F(\sigma)
\exp\big(\I \nu_n(\tau-\sigma)\big)\;.
\end{aligned}
\end{equation}
Since the external force appears only through this action in the exponent of 
the equilibrium density matrix, we may easily evaluate the functional 
derivatives according to (\ref{gli:eq:varqq}). For the position autocorrelation 
function\index{damped harmonic oscillator!position autocorrelation function of}
in imaginary time we thus find
\begin{equation}
C(\tau) = \frac{1}{m\beta}\sum_{n=-\infty}^{\infty}
\frac{\exp(\I \nu_n\tau)}{\nu_n^2+\gamma\vert\nu_n\vert+\omega_0^2}\;.
\label{gli:eq:corrimg}
\end{equation}
Unfortunately, the real time correlation function cannot be obtained simply
by replacing the imaginary time $\tau$ by $\I t$ where $t$ is a real time.
For negative times $t$, the sum in (\ref{gli:eq:corrimg}) would not converge.
We therefore have to perform an analytic continuation to real times in a more
elaborate way.

\begin{figure}[t]
\begin{center}
\includegraphics[width=0.8\textwidth]{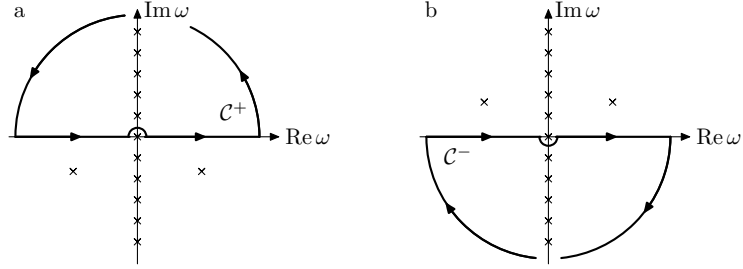}
\end{center}
\caption{The analytic continuation of the imaginary time correlation function
$\langle q(\tau)q(0)\rangle$ to real times makes use of the integration 
contours depicted in (\textbf{a}) and (\textbf{b}) to obtain 
(\protect\ref{gli:eq:sum1}) and (\protect\ref{gli:eq:sum2}), respectively}
\label{gli:fig:contour}
\end{figure}

The idea is to express the sum (\ref{gli:eq:corrimg}) as a contour integral in
the complex frequency plane. To this end, we need a function with poles at 
frequencies $\omega = \I \nu_n, n=-\infty,\dots,\infty$ with a residuum of
one. This requirement is satisfied by $\hbar\beta/[1-\exp(-\hbar\beta\omega)]$.
By integration along the contour shown in
Fig.~\ref{gli:fig:contour}(\textbf{a}) we find
\begin{equation}
\begin{aligned}
&\int_{\cal C^+}\D \omega \frac{\hbar\beta}
{1-\exp(-\hbar\beta\omega)}\frac{\exp(-\omega\tau)}{-\omega^2
+\I \gamma\omega+\omega_0^2}\\
&\hspace{4truecm}= -\I \frac{\pi}{\omega_0^2}
-2\pi\I \sum_{n=1}^{\infty}\frac{\exp({\rm i}\nu_n\tau)}
{\nu_n^2+\gamma\nu_n+\omega_0^2}\;.
\end{aligned}
\label{gli:eq:sum1}
\end{equation}
Similarly, an integration along the contour shown in 
Fig.~\ref{gli:fig:contour}(\textbf{b}) leads to
\begin{equation}
\begin{aligned}
&\int_{\cal C^-}\D \omega \frac{\hbar\beta}
{1-\exp(-\hbar\beta\omega)}\frac{\exp(-\omega\tau)}{-\omega^2
-\I \gamma\omega+\omega_0^2}\\
&\hspace{4truecm}= \I \frac{\pi}{\omega_0^2}
+2\pi\I \sum_{n=-\infty}^{-1}\frac{\exp({\rm i}\nu_n\tau)}
{\nu_n^2+\gamma\nu_n+\omega_0^2}
\end{aligned}
\label{gli:eq:sum2}
\end{equation}
Subtracting (\ref{gli:eq:sum2}) from (\ref{gli:eq:sum1}), the imaginary time
correlation function can be expressed as
\begin{equation}
\begin{aligned}
&\frac{1}{m\beta}\sum_{n=-\infty}^{\infty}\frac{\exp(\I \nu_n\tau)}
{\nu_n^2+\gamma\nu_n+\omega_0^2}\\
&\hspace{2truecm}= \frac{\hbar}{m\pi}\int_{-\infty}^{\infty}
\D \omega\frac{\gamma\omega}{(\omega^2-\omega_0^2)^2+\gamma^2\omega^2}
\frac{\exp(-\omega\tau)}{1-\exp(-\hbar\beta\omega)}\;.
\end{aligned}
\end{equation}
Now we may pass to real times by the replacement $\tau\rightarrow\I t$
to obtain the real time correlation function
\begin{equation}
C(t) = \frac{\hbar}{m\pi}\int_{-\infty}^{\infty}
\D \omega\frac{\gamma\omega}{(\omega^2-\omega_0^2)^2+\gamma^2\omega^2}
\frac{\exp(-\I \omega t)}{1-\exp(-\hbar\beta\omega)}\;.
\label{gli:eq:corrreal}
\end{equation}

Physical insight into this result can be gained by considering the Fourier
transform of this correlation function
\begin{equation}
\tilde C(\omega) = \int_{-\infty}^{\infty}\D t\exp(\I \omega t)C(t)
\label{gli:eq:ftc}
\end{equation}
which may be related to the dynamic susceptibility $\tilde\chi(\omega)$ of the 
damped harmonic oscillator. According to the Ehrenfest theorem the equation of 
motion for the expectation value of the position agrees with the corresponding 
classical equation of motion. In the presence of an external force $F(t)$, the 
latter reads
\begin{equation}
m\ddot q + m\gamma\dot q + m\omega_0^2 q = F(t)\;.
\label{gli:eq:dhoem}
\end{equation}
The solution of this equation may be expressed in terms of the response 
function $\chi(t)$ as
\begin{equation}
q(t) = \int_{-\infty}^t\D s \chi(t-s)F(s)
\end{equation}
which by means of a Fourier transformation becomes
\begin{equation}
\tilde q(\omega) = \tilde\chi(\omega)\tilde F(\omega)\;.
\end{equation}
With the equation of motion (\ref{gli:eq:dhoem}) the dynamic susceptibility
of the damped harmonic oscillator is then found to read
\begin{equation}
\chi(\omega) =\frac{1}{m}\frac{1}{-\omega^2-\I \gamma\omega+\omega_0^2}\;.
\label{gli:eq:dynsusc}
\end{equation}
Together with (\ref{gli:eq:corrreal}) and (\ref{gli:eq:ftc}) one finally 
obtains the relation
\begin{equation}
\tilde C(\omega) = \frac{2\hbar}{1-\exp(-\hbar\beta\omega)}\chi''(\omega)
\label{gli:eq:fdt}
\end{equation}
which represents an example of the so-called fluctuation-dissipation 
theorem\index{fluctuation-dissipation theorem} \cite{gli:calle51}. Here, the 
position autocorrelation function describes the fluctuations while the 
imaginary part of the dynamic susceptibility $\chi''$ can be shown to determine 
the energy dissipated under the influence of the external driving $F(t)$. While 
the relation (\ref{gli:eq:fdt}) is exact for linear systems like the damped 
harmonic oscillator considered here, it still holds for nonlinear systems 
within linear response theory. There, the response to the external force is 
considered in leading order perturbation theory \cite{gli:kubo57,gli:kubo66}.

It is instructive to consider the real time correlation function 
(\ref{gli:eq:corrreal}) in more detail. We first decompose the correlation
function $C(t)$ into its real and imaginary part, or equivalently, into its 
symmetric and antisymmetric part
\begin{equation}
C(t) = S(t)+\I A(t)\;,
\end{equation}
with
\begin{equation}
S(t) = \dfrac{1}{2}\big(\langle q(t)q(0)\rangle + \langle q(0)q(t)\rangle
\big)
\end{equation}
and
\begin{equation}
A(t) = -\dfrac{\I}{2}\big(\langle q(t)q(0)\rangle - \langle q(0)q(t)\rangle
\big)\;.
\end{equation}
From (\ref{gli:eq:corrreal}) we find
\begin{equation}
S(t) = \frac{\hbar}{2\pi m}\int_{-\infty}^{\infty}\D \omega
\frac{\gamma\omega}{(\omega^2-\omega_0^2)^2+\gamma^2\omega^2}\coth\!\left(
\frac{\hbar\beta\omega}{2}\right)\cos(\omega t)
\label{gli:eq:corrsym}
\end{equation}
and
\begin{equation}
A(t) = -\frac{\hbar}{2\pi m}\int_{-\infty}^{\infty}\D \omega
\frac{\gamma\omega}{(\omega^2-\omega_0^2)^2+\gamma^2\omega^2}\sin(\omega t)\;.
\label{gli:eq:corranti}
\end{equation}
Apart from Planck's constant appearing in the prefactor, the antisymmetric part
is purely classical. In fact, one demonstrates within linear response theory 
the general relation
\begin{equation}
\chi(t) = \frac\I {\hbar}\langle [q(t),q(0)]\rangle \uTheta(t) =
-\frac{2}{\hbar}A(t)\uTheta(t)\;.
\end{equation}
Therefore, our statement follows as a consequence of the Ehrenfest theorem
which ensures that the response function $\chi(t)$ of the damped harmonic 
oscillator is classical.

More interesting is the symmetric part (\ref{gli:eq:corrsym}) of the 
correlation function $C(t)$. There exist two different types of time scales 
determined by the poles of the integrand in (\ref{gli:eq:corrsym}). One set of 
four poles at frequencies
\begin{equation}
\omega=\pm\left(\bar\omega\pm\I \frac{\gamma}{2}\right)
\label{gli:eq:dhofreq}
\end{equation}
corresponds to the characteristic frequencies of a damped harmonic oscillator
with the oscillation frequency
\begin{equation}
\bar\omega = \left(\omega_0^2-\frac{\gamma^2}{4}\right)^{1/2}
\end{equation}
shifted by the damping. In addition, there exists an infinite number of poles
at imaginary frequencies $\I \nu_n, n=-\infty,\dots,\infty$ depending on the
temperature of the heat bath via the Matsubara frequencies defined in 
(\ref{gli:eq:matsubara}). With these poles, one can evaluate the integral 
(\ref{gli:eq:corrsym}) by means of a contour integration to obtain
\begin{equation}
\begin{split}
S(t) &= \frac{\hbar}{2m\bar\omega}\exp(-\gamma\vert t\vert/2)
\frac{[\sinh(\hbar\beta\bar\omega)\cos(\bar\omega t)+\sin(\hbar\beta\gamma/2)
\sin(\bar\omega\vert t\vert)]}{\cosh(\hbar\beta\bar\omega)-
\cos(\hbar\beta\gamma/2)}\\
&\quad-\frac{2\gamma}{m\beta}\sum_{n=1}^{\infty}
\frac{\nu_n\exp(-\nu_n\vert t\vert)}
{(\nu_n^2+\omega_0^2)^2-\gamma^2\nu_n^2}\;.
\end{split}
\label{gli:eq:sqq}
\end{equation}
The sum in the second line becomes important at low temperatures 
$k_{\rm B}T\ll \hbar\gamma/4\pi$. In order to discuss this quantum effect, we 
focus in the following discussion on the case of zero temperature. Then, the 
poles of the hyperbolic cotangent in (\ref{gli:eq:corrsym}) become dense and 
form a cut on the imaginary axis which has consequences for the long-time 
behaviour of the correlation function. In the limit $\beta\to\infty$ the sum in 
the second line of (\ref{gli:eq:sqq}) turns into an integral. Noting that the 
long-time behaviour is determined by the behaviour of the integrand at small 
arguments we find
\begin{equation}
S(t)\sim -\frac{\hbar\gamma}{\pi m}\int_0^{\infty}\D x
\frac{x\exp(-x\vert t\vert)} {\omega_0^4} = 
-\frac{\hbar\gamma}{\pi m\omega_0^4}\frac{1}{t^2}\;.
\label{gli:eq:sqqasy}
\end{equation}
Instead of the usual exponential decay we thus find an algebraic decay.

In the limit of vanishing damping, the imaginary part of the dynamic 
susceptibility appearing in the integrand of (\ref{gli:eq:corrsym}) turns
into a sum of two delta functions located at $\omega_0$ and $-\omega_0$. 
For weak but finite damping, the delta functions broaden into Lorentzians
\begin{equation}
\frac{\gamma\omega}{(\omega^2-\omega_0^2)^2+\gamma^2\omega^2} =
\frac{\gamma}{4\bar\omega}\left(\frac{1}{(\omega-\bar\omega)^2+\gamma^2/4}-
\frac{1}{(\omega+\bar\omega)^2+\gamma^2/4}\right)
\label{gli:eq:dsdecomp}
\end{equation}
corresponding to the four poles (\ref{gli:eq:dhofreq}) and one can assume that 
only the integrand in the neighbourhood of these poles is relevant. Within the 
so-called Markov approximation\index{Markov approximation}, one then replaces 
the contributions to the integrand which depend only weakly on frequency by 
their values at $\bar\omega$. As we will see, in contrast to 
(\ref{gli:eq:sqqasy}) the correlation function $S(t)$ at zero temperature then 
no longer decays algebraically. It is interesting to discuss the reason for 
this discrepancy.

To this end we go back to the integral representation (\ref{gli:eq:corrsym}) of 
the correlation function $S(t)$. In a first step, we apply the so-called 
rotating wave approximation (RWA)\index{rotating wave approximation} which
consists in neglecting the Lorentzian located at $-\bar\omega$, i.e. the
second term in (\ref{gli:eq:dsdecomp}) \cite{gli:loudo83}. Physically speaking, 
we neglect processes where the system is excited into an energetically higher 
state by loosing energy to the driving force or where the system decays to a 
lower state by absorbing energy. For finite temperatures, we now have
\begin{equation}
S_{\rm RWA}(t) = \frac{\hbar}{8\pi m\bar\omega}\int_{-\infty}^{\infty}
\D \omega\frac{\gamma}{(\omega-\bar\omega)^2+\gamma^2/4}
\coth\!\left(\frac{\hbar \beta\omega}{2}\right)\cos(\omega t)\;.
\label{gli:eq:rwa}
\end{equation}
Within the Markov approximation\index{Markov approximation}, we replace the 
hyperbolic cotangent by its value at $\omega=\bar\omega$. In the limit of zero 
temperature this leads to
\begin{equation}
\begin{split}
S_{\rm RWA, Markov}(t) &= \frac{\hbar}{8\pi m\bar\omega}\int_{-\infty}^{\infty}
\D \omega\frac{\gamma}{(\omega-\bar\omega)^2 +
\gamma^2/4}\cos(\omega t)\\
&= \frac{\hbar}{4m\bar\omega}\cos(\bar\omega t)\exp\!\left(-\frac{\gamma}{2}t
\right)\;.
\end{split}
\label{gli:eq:rwamarkov}
\end{equation}
We thus find an oscillation with a frequency shifted due to the environmental 
coupling and an exponential decay in contrast to the algebraic decay 
(\ref{gli:eq:sqqasy}).

\begin{figure}[t]
\begin{center}
\includegraphics[width=0.6\textwidth]{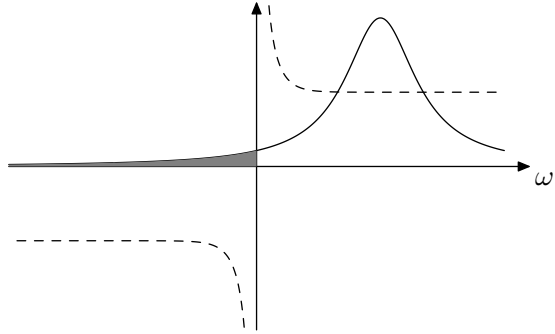}
\end{center}
\caption{The full and dashed lines represent the Lorentzian and hyperbolic
cotangent, respectively, which contribute to the integrand in 
(\protect\ref{gli:eq:rwa})}
\label{gli:fig:markov}
\end{figure}

This difference can be understood by taking a closer look at the Markov 
approximation. In Fig.~\ref{gli:fig:markov} the Lorentzian and the hyperbolic
cotangent appearing in (\ref{gli:eq:rwa}) are schematically shown as full and
dashed line, respectively. In order to obtain (\ref{gli:eq:rwamarkov}), we have 
approximated the hyperbolic cotangent by a constant. However, in doing this, we 
have replaced an antisymmetric function by a symmetric one which can only yield
a non-vanishing result together with the rotating wave approximation made 
above. As a consequence, the area shaded in grey has been taken with the wrong 
sign. The idea was that this difference should be small and arising from 
frequencies far from $\bar\omega$. However, due to the large extent of a 
Lorentzian, it is of order $\gamma$ and, in addition, it replaces an 
exponential decay by an algebraic decay.

At zero temperature the difference between (\ref{gli:eq:rwa}) and 
(\ref{gli:eq:rwamarkov}) becomes
\begin{equation}
\begin{split}
\Delta &= S_{\rm RWA}(t)-S_{\rm RWA, Markov}(t)\\
&= -2\frac{\hbar}{8\pi m}\int_{-\infty}^0\D \omega \frac{\gamma}
{(\omega-\bar\omega)^2+\gamma^2/4}\cos(\omega t)
\end{split}
\end{equation}
which may be expressed in terms of integral sine and integral cosine functions.
For our purpose it is sufficient to note that for long times, the difference
indeed decays algebraically as
\begin{equation}
\Delta = -\frac{\hbar\gamma}{2\pi m\bar\omega_0^4}\frac{1}{t^2}\;.
\end{equation}
We remark that the factor of two relative to the result (\ref{gli:eq:sqqasy})
arises because no rotating wave approximation has been made in deriving the
latter. 

In the previous discussion, we have been concerned with an effect of order
$\gamma$ which in the spirit of the Markov approximation should be considered
as small. However, the Markov approximation changes the long-time behaviour of 
the equilibrium correlation function $S(t)$ qualitatively and the absence of an 
algebraic decay, even with a small prefactor, may be relevant. For truly weak
coupling, the damping constant $\gamma$ should be the smallest frequency
scale.  Apart from the usual weak coupling condition $\gamma\ll\omega_0$ we 
also have to require $\gamma\ll k_{\rm B}T/\hbar$ which at low temperatures may 
become difficult to satisfy.\index{damped harmonic oscillator|)}

\section*{Acknowledgment}
This chapter is based on courses taught at the Max-Planck-Institut f\"ur Physik
komplexer Systeme in Dresden and at the Universit\'e Louis Pasteur Strasbourg.
The author is grateful to the organizers of the school on ``Coherent evolution
in noisy environments'', in particular Andreas Buchleitner (Dresden), as well 
as Rodolfo Jalabert (ULP Strasbourg) for having provided the opportunity to 
lecture on path integrals and quantum dissipation. The participants of the two
courses through their questions contributed to these lecture notes. It is a
pleasure to acknowledge inspiring conversations with Giovanna Morigi which have 
motivated some of the discussions in particular in the last section. 
Furthermore, I am indebted to Peter H{\"a}nggi for several useful comments on 
an earlier version of the manuscript and for drawing my attention to some
less known references.

%

\end{document}